\documentclass[fleqn,usenatbib]{mnras}

\usepackage{newtxtext,newtxmath}
\usepackage[T1]{fontenc}

\DeclareRobustCommand{\VAN}[3]{#2}
\let\VANthebibliography\thebibliography
\def\thebibliography{\DeclareRobustCommand{\VAN}[3]{##3}\VANthebibliography}

\usepackage{graphicx}	% Including figure files
\usepackage{amsmath}	% Advanced maths commands
\usepackage[utf8]{inputenc}
\title[Radiative Compression in the Pillars of Creation]{Radiative Compression of Dense Cores in the Pillars of Creation as Revealed by \emph{JWST} Extinction Mapping}

\author[J. Li et al.]{
Jun Li,$^{1}$\thanks{E-mail: lijun@gzhu.edu.cn (JL)}
Bingqiu Chen,$^{2}$\thanks{E-mail: bchen@ynu.edu.cn (BQC)}
He Zhao,$^{3}$\thanks{E-mail: he.zhao@oca.eu (HZ)}
Jian Gao,$^{4,5}$
Xi Chen$^{1}$
\\
$^{1}$Center for Astrophysics, Guangzhou University, Guangzhou 510006, People's Republic of China\\
$^{2}$South-Western Institute for Astronomy Research, Yunnan University, Kunming, Yunnan 650091, People's Republic of China\\
$^{3}$Departamento de Fisica y Astronomia, Facultad de Ciencias Exactas, Universidad Andres Bello, Fernandez Concha 700, 8320000 Santiago, Chile\\
$^{4}$Institute for Frontiers in Astronomy and Astrophysics, Beijing Normal University, Beijing 102206,  People's Republic of China\\
$^{5}$Department of Astronomy, Beijing Normal University, Beijing 100875, People's Republic of China
}

\date{Accepted 2026 March 20. Received 2026 March 15; in original form 2025 October 30}

\pubyear{\the\year{}}

\begin{document}
\label{firstpage}
\pagerange{\pageref{firstpage}--\pageref{lastpage}}
\maketitle

\begin{abstract}
The Pillars of Creation in M16 represent an iconic star-forming region where stellar feedback shapes molecular cloud evolution. We present a detailed investigation of dust extinction and density structure in the Pillars of Creation using multiband photometric observations from \emph{JWST} NIRCam. A high-resolution (2\arcsec) extinction map reaching depths of $A_V\sim 100$ mag has been constructed using NIRCam filters F090W, F200W, F335M, and F444W. This map clearly reveals the intricate structure of dense gas within the molecular cloud in the Pillars of Creation region. Analysis of the column density probability distribution function (N-PDF) exhibits a characteristic lognormal distribution at intermediate extinctions ($A_V\approx10-30$\,mag), which transitions to a power-law tail at high extinctions ($A_V\gtrsim$ 30\,mag) where star-forming cores reside. The power-law slope $\alpha$ displays significant spatial variation, steepening from $\alpha\approx 2.0$ at the pillar tips facing the NGC 6611 cluster to $\alpha\approx$4.0 in regions distant from the cluster. This systematic gradient demonstrates that stellar feedback not only disperses molecular clouds but can also locally enhance the formation of dense, self-gravitating structures through radiative compression.
\end{abstract}


%%%%%%%%%%%%%%%%%%%%%%%%%%%%%%%%%%%%%%%%%%%%%%%%%%

%%%%%%%%%%%%%%%%% BODY OF PAPER %%%%%%%%%%%%%%%%%%

\section{Introduction}

Understanding the internal structure of molecular clouds is fundamental to unraveling star formation mechanisms. Regions experiencing strong stellar feedback from radiation and winds of massive stars present particularly valuable laboratories, as these environments demonstrate how feedback simultaneously disrupts clouds while potentially triggering new star formation through compression \citep{Elmegreen1998, Gritschneder2009, Walch2012}. High-resolution studies of such regions are crucial for identifying the physical processes governing gas fragmentation and core collapse.
			
The Eagle Nebula (M16) is a well-known star-forming complex located in the constellation Serpens, at a distance of approximately $1.74 \pm 0.13$\,kpc \citep{Kuhn2019}. The region is illuminated and shaped by the young stellar cluster NGC~6611, which hosts several massive stars, including four early-type O stars, with an estimated age of 2 -- 3\,Myr \citep{Dufton2006}. Ultraviolet radiation from these stars photoionizes and heats the surrounding gas, carving out prominent features such as the ``Pillars of Creation''—dense columns of gas extending toward the cluster. These structures were famously imaged in H$\alpha$ by Hubble Space Telescope (\emph{HST}) \citep{Hester1996} and have since been studied across multiple wavelengths. Submillimeter and molecular line observations have revealed dense cores at the tips of the pillars \citep[e.g.,][]{White1999}, while X-ray data from \textit{Chandra} have identified embedded young stellar objects (YSOs) within the pillars \citep{Linsky2007}. Far-infrared observations from \emph{Herschel} Space Observatory further show that the influence of NGC~6611 extends throughout the cloud, heating both the pillars and the surrounding molecular gas \citep[e.g.,][]{Hill2012}. Recent James Webb Space Telescope (\emph{JWST}) observations of the Pillars of Creation have identified 253 young stellar object candidates that show spatial correlations with feedback-driven structures and age gradients suggestive of triggered star formation \citep{Wen2025}. These multiwavelength studies establish M16 as an ideal laboratory for investigating how stellar feedback shapes dense gas and regulates star formation \citep{Guarcello2010, Indebetouw2007}.
			
Extinction mapping provides one of the most direct methods for probing the internal structure molecular cloud. Traditional near-infrared (NIR) techniques based on ground-based observations suffer from saturation at moderate column densities ($A_V \lesssim 30$\,mag), rendering them ineffective for studying the densest cores where stars form \citep{McCaughrean2002, Sugitani2002}. Even mid-infrared (MIR) observations from \emph{Spitzer} reach only up to $A_V \sim 50-$60\,mag in the most opaque regions \citep{Indebetouw2007, Kainulainen2013}. Column densities inferred from \emph{Herschel} far-infrared emission reach up to $\sim 10^{23}$\,cm$^{-2}$ \citep{Hill2012, Tremblin2013}, but the angular resolution of $\sim 36$\,arcsec smooths over fine substructures. The launch of the \emph{JWST}, with its unprecedented sensitivity and resolution in the NIR and MIR, now enables much more detailed extinction mapping of dense molecular cloud. 
			
Characterizing the density structure of molecular clouds is critical for understanding the relative roles of turbulence and gravity in star formation. In a turbulent medium, the column density probability distribution function (N-PDF) is expected to follow a log-normal form, arising from the multiplicative nature of density fluctuations \citep[e.g.,][]{Vazquez1994, Hill2008, Kainulainen2013}. When self-gravity becomes dominant, a high-density power-law tail emerges in the N-PDF \citep[e.g.,][]{Shu1977, Klessen2000, Federrath2013}. Observations of nearby clouds confirm this pattern: non-star-forming regions typically show log-normal N-PDFs, while active star-forming regions exhibit additional power-law tails at high column densities \citep{Kainulainen2009, Kainulainen2011, Schneider2013, Schneider2022}. Furthermore, the slope of the power-law tail correlates with star formation activity—regions with flatter slopes tend to host a larger fraction of young protostars \citep{Stutz2015}. Thus, the N-PDF serves as a powerful diagnostic linking cloud structure to star formation.

\emph{JWST} has transformed molecular cloud studies by enabling high-resolution extinction mapping in regions of extreme reddening, allowing investigation of small-scale structure in star-forming clouds. Recently, NIRCam observations have been instrumental in precisely constraining NIR and MIR extinction laws across diverse environments, from local star-forming regions like the Pillars of Creation \citep{Li2024} and the Galactic Center \citep{Bravo2025}, to extragalactic domains such as 30 Doradus \citep{Fahrion2023}. In addition, \emph{JWST} photometry has been utilized to construct highly detailed extinction maps, revealing dense filamentary structures and widespread ice absorption (e.g., CO and H$_2$O ices) in the Milky Way clouds \citep{Ginsburg2023,Gramze2025}. Furthermore, this high-resolution stellar extinction technique has even been successfully extended to map the cold dust distribution in the circumnuclear disks of nearby active galaxies \citep{Vermot2025}. 

In this work, we utilize multi-band \emph{JWST} NIRCam photometry to construct an ultradeep, high-resolution extinction map of the Pillars of Creation in M16. We analyze the resulting structure to investigate the role of dense gas in ongoing star formation. This paper is organized as follows: Section~\ref{sec:data} describes the \emph{JWST} data and method; Section~\ref{sec:res} presents the results and discussions; Section~\ref{sec:sum} summarizes our findings and conclusions.
 
%\emph{JWST} has transformed molecular cloud studies by enabling high-resolution extinction mapping in regions of extreme reddening, allowing unprecedented investigation of small-scale structure in star-forming clouds. In this work, we utilize multi-band \emph{JWST} NIRCam photometry to construct an ultradeep, high-resolution extinction map of the Pillars of Creation in M16. We analyze the resulting structure to investigate the role of dense gas in ongoing star formation. This paper is organized as follows: Section~\ref{sec:data} describes the \emph{JWST} data and method; Section~\ref{sec:res} presents the results and discussions; Section~\ref{sec:sum} summarizes our findings and conclusions.

\section{Data and method} \label{sec:data}
			
\begin{figure*}
	\includegraphics[scale=0.5]{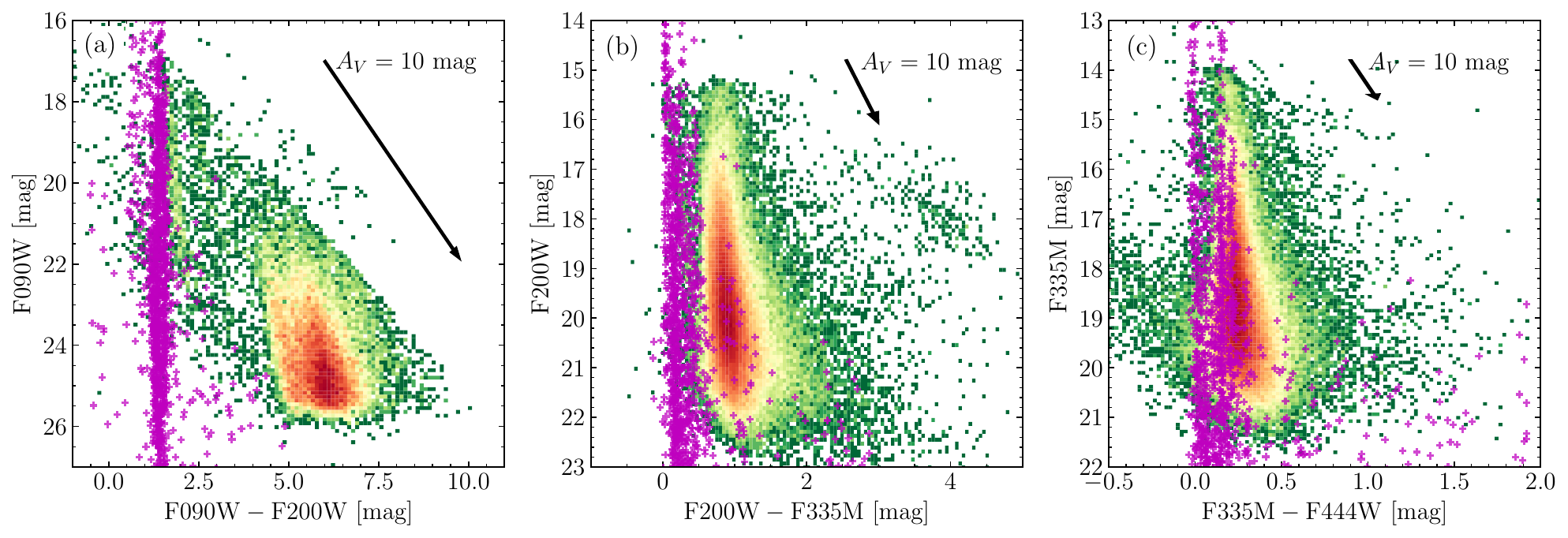}
	\caption{Color-magnitude diagrams derived from \emph{JWST}/NIRCam photometry of sources detected in the Pillars of Creation region in M16. The panels show F090W versus F090W$-$F200W (a), F200W versus F200W$-$F335M (b), and F335M versus F444W$-$F335M (c). Magenta points indicate synthetic stellar populations generated using the TRILEGAL Galactic model for the same field of view. Black arrows in each panel shows the reddening vector of $A_V=10$ mag calculated using the extinction law of \protect\cite{Hensley2023}.}
	\label{fig:ccds}
\end{figure*}

We use \emph{JWST}/NIRCam imaging of the Eagle Nebula (M16) obtained under the Director's Discretionary program (Proposal ID: 2739; PI: Pontoppidan, Klaus M.). These observations cover the famous Pillars of Creation region and were previously presented by \cite{Li2024}, who combined the NIRCam with \emph{JWST}/MIRI data to derive the dust extinction law from 0.9$-$7.7\,$\mu$m. The calibrated images are publicly available from the Mikulski Archive for Space Telescopes (MAST). In this work we adopt the NIRCam source catalogs produced by \cite{Li2024}, which were generated using PSF-fitting photometry with the \emph{JWST}-specific \textsc{StarbugII} pipeline \citep{Nally2023}. We apply a quality cut requiring that the magnitude uncertainty in each band less than 0.5 mag. After this cut, the catalog contains $\sim2\times10^4$ sources detected in the short-wavelength filters (F090W and F200W) and $\sim10^5$ sources in the longer-wavelength set (F200W, F335M, F444W). This strong wavelength dependence arises because the Pillars have extremely high extinction in F090W band, so far fewer background stars are visible in F090W band. In contrast, the longer-wavelength bands suffer less dust extinction and recover more sources. The color-magnitude distributions of the filtered catalog are shown in Figure~\ref{fig:ccds}. 

Most of the stars detected toward the Pillars of Creation are background objects lying behind the dust cloud, so the Near-Infrared Color Excess (NICE; \citealt{Lada1994}) method can be applied to measure extinction. To minimize contamination by foreground stars, we apply a color cut based on the F090W vs.\ F090W$-$F200W color-magnitude diagram in Figure \ref{fig:ccds}(a). Given that the foreground extinction toward M16 to be $A_V\approx 3$\,mag \citep{Hillenbrand1993}, we exclude sources with $(\mathrm{F090W}-\mathrm{F200W})<3.5$\,mag, which removes most foreground field stars. For the remaining stars, the color excess between two NIRCam bands $\lambda_1$ and $\lambda_2$ is defined as
\begin{equation}
E(m_{\lambda_1}-m_{\lambda_2})=(m_{\lambda_1}-m_{\lambda_2})_{\rm obs}-(m_{\lambda_1}-m_{\lambda_2})_0
\end{equation}
where $(m_{\lambda_1}-m_{\lambda_2})_{\rm obs}$ is the observed colors and $(m_{\lambda_1}-m_{\lambda_2})_0$ denotes the intrinsic colors without reddening.  The color excess $E$ is directly proportional to the absolute extinction (e.g. visual extinction $A_V$) or dust column density.
			
We compute color excesses using \emph{JWST}/NIRCam photometry in four bands, forming three combinations: $E({\rm F090W}-{\rm F200W})$, $E({\rm F200W}-{\rm F335M})$, and $E({\rm F335M}-{\rm F444W})$.  To estimate the intrinsic colors, we generate a synthetic stellar population toward M16 with the TRILEGAL\footnote{http://stev.oapd.inaf.it/cgi-bin/trilegal} Galactic model \citep{Girardi2005}, processed through the \emph{JWST} filter curves. The synthetic stars are plotted in Figure \ref{fig:ccds} by magenta points. We adopt the mean of each simulated color distribution as $(m_{\lambda_1}-m_{\lambda_2})_0$ and its standard deviation as the corresponding uncertainty.  This yields intrinsic colors $\rm (F090W-F200W)_0 = 1.37 \pm 0.28$\,mag, $\rm (F200W-F335M)_0 = 0.21 \pm 0.14$\,mag, and $\rm (F335M-F444W)_0 = 0.11 \pm 0.08$\,mag, respectively.
			
Not all sources are detected in all four bands, particularly in F090W suffers the highest extinction. To maximize sample size, we employ a hybrid approach.  We first compute $E(\mathrm{F200W}-\mathrm{F335M})$ for stars detected in both bands.  For stars lacking one of those bands, we compute either $E(\mathrm{F090W}-\mathrm{F200W})$ or $E(\mathrm{F335M}-\mathrm{F444W})$ (whichever colors are available) and convert to the $E(\mathrm{F200W}-\mathrm{F335M})$ basis using reddening ratios measured in M16 by \citet{Li2024}: $E(\mathrm{F200W}-\mathrm{F335M})/E(\mathrm{F090W}-\mathrm{F200W})= 0.123$ and $E(\mathrm{F335M}-\mathrm{F444W})/E(\mathrm{F200W}-\mathrm{F335M})=0.280$.  In practice, our final sample contains $\sim 9.5\times10^4$ stars with direct $E({\rm F200W}-{\rm F335M})$ measurements, plus $\sim1.4\times10^3$ and $\sim3.7\times10^3$ stars with excesses recovered via the $E(\mathrm{F090W}-\mathrm{F200W})$ or $E(\mathrm{F335M}-\mathrm{F444W})$ conversions, respectively.
			
%The color excesses are then converted to visual extinction $A_V$ by using the reddening law of the astrodust+PAH model from \citet{Hensley2023}, which gives $A_V/E(\mathrm{F200W}-\mathrm{F335M})=26.7$. To create a continuous extinction map, we then spatially smooth the stellar extinctions. Specifically, we convolve the discrete $A_V$ values with a Gaussian kernel (FWHM = 2\arcsec, pixel scale = 1\arcsec), computing for each map pixel the weighted average of stars within 3$\times$FWHM ($6''$). 

The color excesses are then converted to visual extinction $A_V$ by using the reddening law of the astrodust+PAH model from \citet{Hensley2023}, which gives $A_V/E(\mathrm{F200W}-\mathrm{F335M})=26.7$. The spatial distribution of these field stars, color-coded by their derived $A_V$ values, is presented by Figure \ref{fig:appendix}  in Appendix \ref{sec:appendixA}. To create a continuous extinction map, we then spatially smooth the stellar extinctions. Specifically, we convolve the discrete $A_V$ values with a Gaussian kernel (FWHM = 2\,\arcsec, pixel scale = 1\,\arcsec), computing for each map pixel the weighted average of stars within 3$\times$FWHM ($6''$).  To evaluate the spatial sampling of our field stars, we calculated the nearest-neighbour distances for this final sample of $\sim 10^5$ stars. The mean nearest-neighbour distance is 0.57\,\arcsec, with a median of  0.51\,\arcsec\ and a minimum of 0.02\,\arcsec. Because the typical separation between probing stars ($\sim$0.5\,\arcsec) is sufficiently smaller than the 2\,\arcsec\ FWHM of our Gaussian kernel, our map is well-sampled. This ensures that the resulting continuous map effectively resolves the true physical structures of the cloud without suffering from severe interpolation artifacts or over-smoothing.

\begin{figure*}
	\centering
	\includegraphics[width=0.7\textwidth]{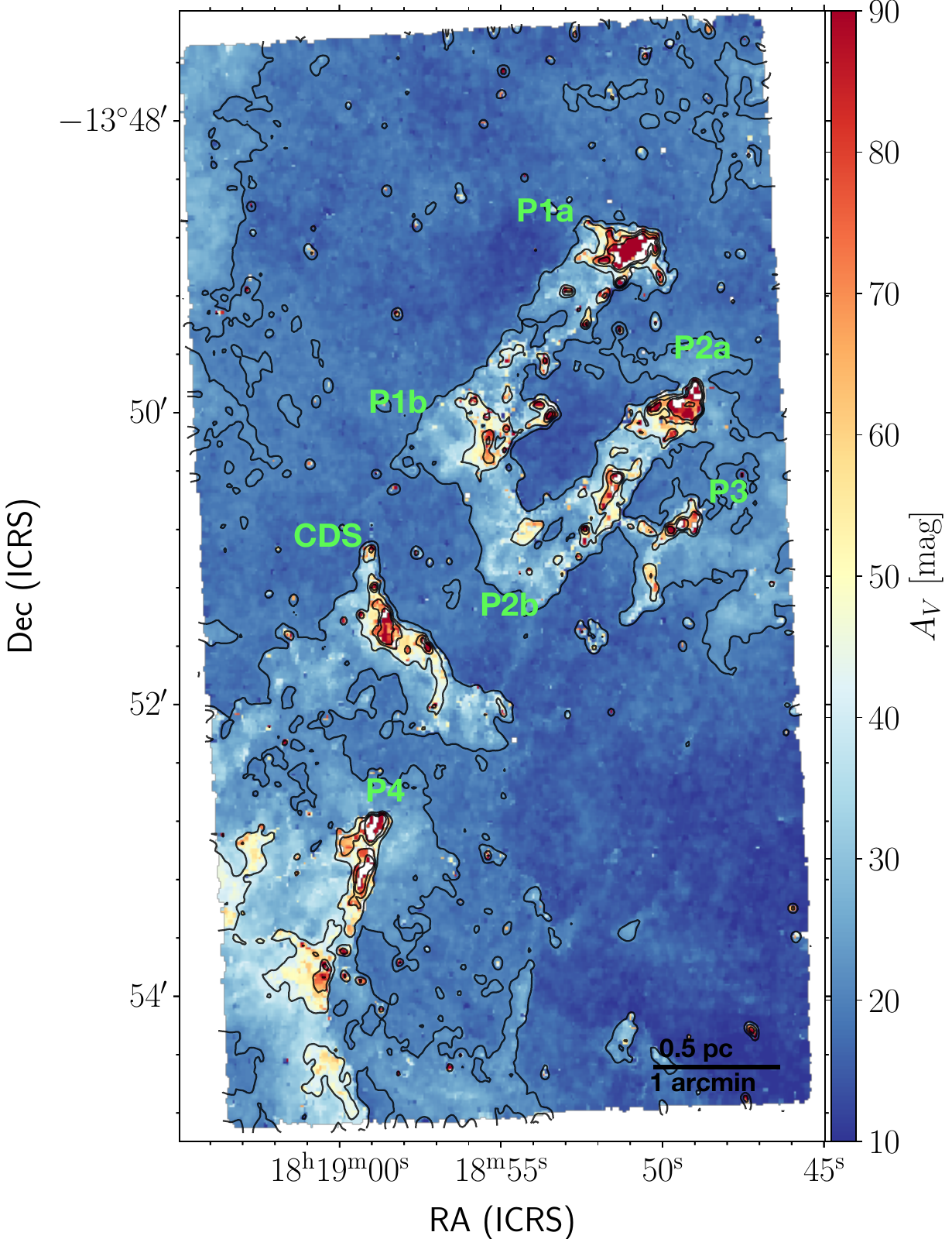}
	\caption{Extinction ($A_V$) map of the Pillars of Creation in M16, derived from \emph{JWST}/NIRCam photometry.  The map has pixel size of 1\arcsec\ and is smoothed with a Gaussian kernel of FWHM = 2\arcsec.  The contours are drawn in black at $A_V=$ [20, 40, 60, 80] mag. A scale bar of 0.5 pc ($\sim$ 1 arcmin) is shown in the lower right corner. The seven labeled subregions (P1a, P1b, P2a, P2b, P3, P4, and the Central Dense Structure, CDS) follow the nomenclature of \protect\cite{Dewangan2024} and \protect\cite{Karim2023}. }
	\label{fig:ce_map}
\end{figure*}

\section{Results and Discussions} \label{sec:res}

\subsection{Extinction Map}

The resulting extinction map is shown in Figure~\ref{fig:ce_map}. For reference, we label seven subregions following the nomenclature of \cite{Dewangan2024} and \cite{Karim2023}: P1a, P1b, P2a, P2b, P3, P4, and the Central Dense Structure (CDS). Our extinction map reveals the iconic pillar structures of M16 with unprecedented detail, which exhibits a highly structured morphology dominated by four main pillar features (labeled P1a/b to P4) that extend from the southeast toward the northwest, consistent with photoevaporative sculpting by the nearby NGC 6611 cluster. The background extinction across the field varies between $A_V\sim 10-$20\,mag, representing the less dense molecular cloud component, while the pillar structures show dramatically enhanced extinction with values reaching $A_V\sim 50-$70\,mag along their main ridges. The densest regions, particularly within the heads of pillars P1a and P2a, exhibit extreme extinction values approaching $A_V>100$\,mag. The sharp extinction gradients are evident at the pillar boundaries, with $A_V$ dropping from $>50$\,mag to background levels over distances within $\lesssim 2''$, consistent with photodissociation region interfaces shaped by UV radiation from the adjacent OB association \citep{Pound1998,Schuller2006}. The pillar orientations and morphologies, with their characteristic cometary shapes and bright-rimmed edges facing the ionizing sources to the northwest, provide clear evidence of ongoing stellar feedback processes.

\subsection{Probability Distribution Functions of Column Density}

\begin{figure*}
	\centering
	\includegraphics[width=1\textwidth]{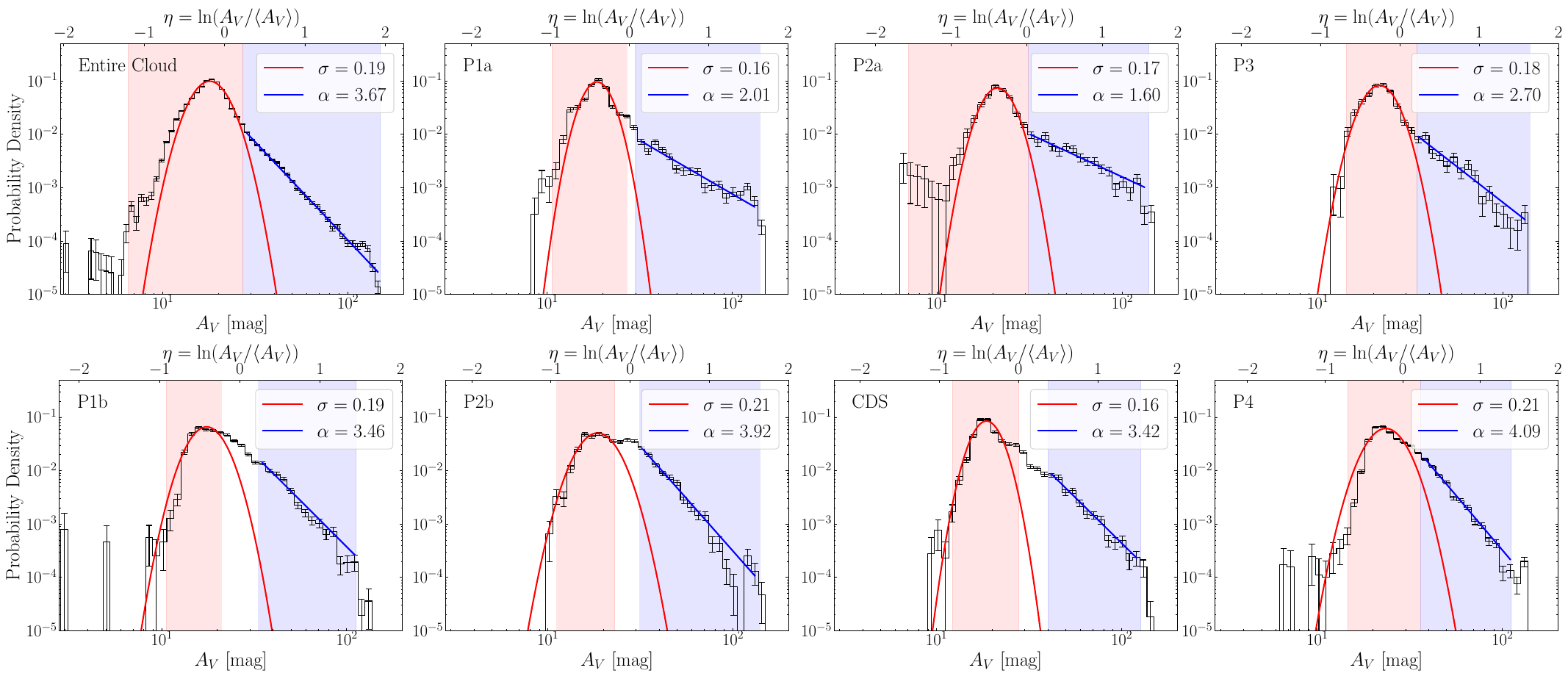}
	\caption{Column-density probability density functions (N-PDFs) derived from our extinction map for the entire Pillars of Creation region and each of the seven subregions (panels labeled as in Figure \ref{fig:ce_map}).  The lower $x$-axis shows the visual extinction $A_V$, the upper $x$-axis gives the corresponding  $\eta = {\rm ln}(A_V/\langle A_V\rangle)$. The $y$-axis is the normalized probability density $p(\eta)$.  Histogram with Poisson error bars shown in each bin. Red curves are the best-fit lognormal functions to the low- and moderate-density ranges, and blue lines are the best-fit power-law tails at high density. The specific fitting ranges for the log-normal and power-law components are indicated by the red and blue shaded regions, respectively. The lognormal widths $\sigma$ and power-law indices $\alpha$ are indicated in each panel.}
	\label{fig:npdf_subregions}
\end{figure*}

The upper left panel of Figure~\ref{fig:npdf_subregions} shows the column density probability distribution function (N-PDF) for the entire Pillars of Creation region, fitted with a lognormal component plus a high-density power-law tail. The fitting procedure is detailed in Appendix \ref{sec:npdf_fit}, with best-fit parameters listed in Table \ref{tab:fit_res}. For the entire Pillars of Creation region, the best-fit lognormal width is $\sigma=0.19$, and the high-extinction tail has a slope of $\alpha=3.67$. This shape is broadly similar to other molecular clouds: the lognormal portion reflects the turbulent density fluctuations with moderate dispersion, and the steep tail indicates self-gravitating gas at high densities \citep{Chen2018}. The power-law slope ($\alpha$ = 3.67) in the high-density regime is relatively steep for typical values ($\alpha\approx1-$4) observed in active star-forming regions \citep[e.g.][]{Kainulainen2009,Schneider2013,Schneider2022}, suggesting that gravity has not yet dominated the cloud structure and that the dense gas fraction remains relatively low in the Pillars of Creation.

Our high-resolution extinction map allows us to explore the spatial variation of the N-PDF by dividing it into subregions (as defined in Figure~\ref{fig:appendix}). The results of fitting the N-PDFs in these subregions are also presented in Figure~\ref{fig:npdf_subregions} and Table~\ref{tab:fit_res}. We find that the lognormal width $\sigma$ is nearly constant across all subregions ($\sigma\sim0.16-$0.21).  The width $\sigma$ of the lognormal distribution is directly related to the turbulent Mach number $M$ through $\sigma=A\,{\rm ln}(1+b^2M^2)^{1/2}$, where $A$=0.11 is a scaling constant from volumn density to column density, and $b\approx0.3-$0.5 is the turbulence forcing parameter  \citep{Burkhart2012,Burkhart2017}.  The implied Mach number is $M\approx1.0-$2.4, indicating mildly supersonic turbulence throughout the cloud. In other words, the spread of intermediate column densities does not vary significantly from one subregion to another. This suggests similar turbulent Mach numbers or forcing parameters in all subregions \citep{Federrath2013,Padoan1997}.  In practice, the modest variation in $\sigma$ indicates that the physics—such as turbulent driving mechanism and thermal state—are comparable throughout M16.

However, the power-law slope shows a pronounced spatial gradient. The three subregions at pillar tips nearest NGC 6611 (P1a, P2a, and P3) display flat high-density tails with $\alpha \approx 2.0$ (Table~\ref{tab:fit_res}), while the four more distant subregions (P1b, P2b, CDS, and P4) show steeper tails with $\alpha \approx 3.4-$4.1. A flatter tail with smaller $\alpha$ indicates a relatively large fraction of very dense gas, whereas a steep tail signals that dense gas is rarer. This behavior is consistent with both theory and observations of star-forming regions. Simulations show that under pure turbulence the N-PDF is lognormal, but once gravity becomes important a power-law tail develops and becomes flatter as collapse proceeds \citep{Klessen2000,Ward2014,Federrath2013}. Observationally, active star-forming clouds (with many young protostars) tend to have shallow N-PDF tails, whereas quiescent clouds show steeper tails \citep{Kainulainen2009,Schneider2013,Stutz2015}. For example, \cite{Stutz2015} found that regions of Orion A with a higher fraction of Class 0 protostars have flatter N-PDF tails, suggesting that a shallow slope is a signature of recent or ongoing star formation. 

We interpret the spatial gradient of $\alpha$ (from $\sim4.0$ far from the cluster to $\sim2.0$ near the cluster) as a clear imprint of feedback from NGC 6611. The intense UV radiation and stellar winds from the OB cluster drive ionization fronts into the surrounding gas, which preferentially removes low-density gas material and compresses the pillar heads \citep[``radiation-driven implosion";][]{Bisbas2011} . Numerical studies show that an expanding H{\sc ii} region creates a compressed shell of dense gas, which flattens the high-density N-PDF tail \citep{Tremblin2013,Tremblin2014}. Such irradiated condensations often have steep compressed radial profiles, sometimes recognizable in the flattening of the power-law tail \citep{Tremblin2014}. Our pillar-tip regions (P1a, P2a, P3) match this scenario: feedback has boosted the fraction of very high column densities at the illuminated edges of the pillars, yielding a flat tail ($\alpha\approx2.0$), and may trigger gravitational instability in marginally stable clumps, converting them into bound, collapsing cores. However, in the far-side subregions, where external compression is weaker, turbulence and self-gravity have been less driven, so the N-PDF tail remains steep ($\alpha\sim3.4-$4.1), as expected for more quiescent gas \citep{Kainulainen2009,Ward2014}

In summary, the N-PDF fits reveal that the dense gas structure in M16 is not uniform: the irradiated pillar heads host an unusually large fraction of dense, self-gravitating gas (hence the flat tail), while shielded regions have a more turbulent distribution. This finding supports the view that stellar feedback from NGC 6611 is actively compressing the pillars and accelerating star formation there \citep[see also, e.g.,][]{Kainulainen2014,Stutz2015}. The systematic flattening of the high-density tail toward NGC 6611 is a striking observational signature of feedback-regulated star formation \citep{Bisbas2011,Tremblin2014}. The spatial variation in N-PDF slopes may provide direct observational evidence for the dual nature of stellar feedback in regulating star formation. While feedback is often invoked as a mechanism for quenching star formation through cloud dispersal, our results demonstrate that it can locally enhance star formation by creating overdense, gravitationally unstable structures. The observed variation in the N-PDF slope across different subregions, which reflects their relative exposure to feedback from NGC 6611, provides a useful diagnostic tool for probing feedback efficiency.
%The quantitative relationship between feedback intensity (parameterized by distance from NGC 6611) and N-PDF slope provides a new diagnostic tool for measuring feedback efficiency.

%\begin{table*}
%	\centering
%	\caption{Parameters of the fits to the N-PDFs in Figure~\ref{fig:npdf_subregions}.}
%	\label{tab:fit_res}
%	\begin{tabular}{c c c c c  c }
%		\hline \hline
%		Regions &   $\langle A_V\rangle$ & Log-normal range&   Power-law tail range & $\sigma$ & $\alpha$ \\
%		& (mag)  & (mag)  &(mag) & & \\
%		\hline
%		Entire Cloud   & 21.6  & [5,30] & [35,150] & 0.19 & 3.47 \\
%		P1a   & 28.8  & [10,30] & [35,150]   & 0.17 & 2.04  \\
%		P2a   & 35.4 & [12,28] & [35,150] & 0.17 & 1.62   \\
%		P3   & 28.6 & [12,30]  & [35,140] & 0.18 & 2.69   \\
%		P1b   & 26.4 & [10,20] & [25,140]  & 0.16 & 3.16  \\
%		P2b   & 28.9  & [10,20] & [32,140] & 0.17 & 3.92   \\
%		CDS  & 28.7  & [10,30] & [40,150]  & 0.17 & 3.42  \\
%		P4   & 31.5  & [12,28]  & [35,140] &  0.15 & 3.81  \\
%		\hline
%	\end{tabular}
%\end{table*}

\begin{table*}
	\centering
	\caption{Parameters of the fits to the N-PDFs for various subregions in Figure~\ref{fig:npdf_subregions} and the entire cloud across different tracers in Figure \ref{fig:comparison}.}
	\label{tab:fit_res}
	\begin{tabular}{l c c c c c}
		\hline \hline
		Region / Tracer & Mean Value & Log-normal range & Power-law tail range & \(\sigma\) & \(\alpha\) \\
		\hline
		\multicolumn{6}{c}{Dust Extinction (\(A_V\) in mag)} \\
		\hline
		Entire Cloud & 21.6 & [7, 27]  & [27, 151] & 0.19 & 3.67 \\
		P1a          & 28.8 & [11, 27] & [30, 141] & 0.16 & 2.01 \\
		P2a          & 35.4 & [7, 31] & [31, 140] & 0.17 & 1.60 \\
		P3           & 28.6 & [14, 34] & [34, 140] & 0.18 & 2.70 \\
		P1b          & 26.4 & [11, 21] & [33, 113] & 0.19 & 3.46 \\
		P2b          & 28.9 & [11, 23] & [32, 140] & 0.21 & 3.92 \\
		CDS          & 28.7 & [12, 28] & [40, 127] & 0.16 & 3.42 \\
		P4           & 31.5 & [15, 36] & [36, 111] & 0.21 & 4.09 \\
		\hline
		\multicolumn{6}{c}{Dust Coulumn Density (\(N(\mathrm{H}_2)\) in \(\times 10^{20}\,\mathrm{cm}^{-2}\))} \\
		\hline
		Entire Cloud & 78  & [43, 78]   & [78, 313]    & 0.13  & 3.51 \\
		\hline
		\multicolumn{6}{c}{\(^{12}\mathrm{CO}\) (\(W_{12\mathrm{CO}}\) in K km s\(^{-1}\))} \\
		\hline
		Entire Cloud & 47  & [22, 45]   & [45, 108]    & 0.25  & 2.45  \\
		\hline
	\end{tabular}
\end{table*}

\subsection{Comparison with Earlier Observations}

\begin{figure*}
	\centering
	\includegraphics[width=0.9\textwidth]{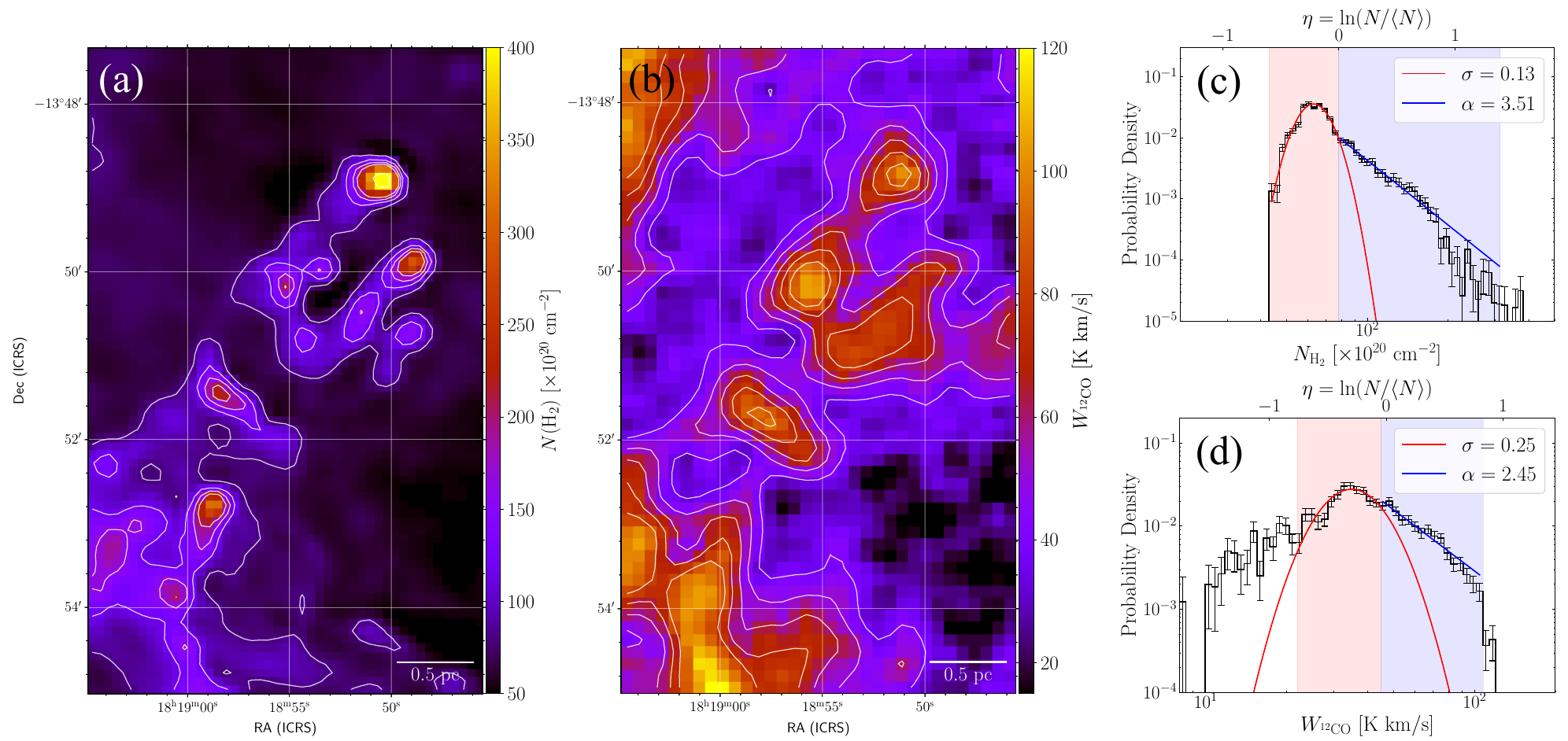}
	\caption{Comparison with previous large-scale observations of M16.  (a) H$_2$ column density map derived from \emph{Herschel} far-infrared dust emission at 36\arcsec\ resolution \citep{Molinari2010}. (b) $\rm ^{12}CO(J=1-0)$ integrated intensity map from the FUGIN survey \citep{Umemoto2017} at $\sim$33\arcsec\ resolution, integrated over 19.3-27.7 km/s. (c) N-PDF for the \emph{Herschel} map (panel a); (d) N-PDF for the $^{12}$CO map (panel b). The best-fit lognormal width $\sigma$ and high-$N$ power-law index $\alpha$ are labeled for each N-PDF, with $\sigma=0.13$, $\alpha=3.51$ for the Herschel N-PDF and $\sigma=0.25$, $\alpha=2.45$ for the $^{12}$CO N-PDF, respectively. The red and blue shaded regions in panels (a) and (b) show the specific fitting ranges for the log-normal and power-law components, respectively. }
	\label{fig:comparison}
\end{figure*}

We have also examined previous observations on the dust or gas distribution in the Pillars of Creation.  Figure~\ref{fig:comparison}(a) shows the $\rm H_2$ column density map derived from Herschel far-IR emission with spatial resolution of $\sim 36''$ \citep{Molinari2010}, which peaks at $N_{\rm H_2}\sim4\times10^{22}\,$cm$^{-2}$ at the tip of P1a.  Figure~\ref{fig:comparison}(b) displays the FUGIN $^{12}$CO(J=1-0) integrated intensity $W_{^{12}\rm CO}$ (at velocity range 19.3-27.7 km/s) with spatial resolution of $\sim 33''$, with values 20$-$100\,$\rm K\, km/s$ \citep{Umemoto2017}.  Using a standard CO-to-H$_2$ conversion factor $X_{\rm CO}\sim 2\times 10^{20}$ \,$\rm cm^{-2}\,(K\,km\,s^{-1})^{-1}$\citep{Bolatto2013,Chen2015}, this corresponds roughly to $N(\mathrm{H}_2)\sim(4-20)\times10^{21}\,$cm$^{-2}$.  Both maps reveal the same large-scale pillar morphology seen in our extinction map in Figure \ref{fig:ce_map}, with enhanced column densities tracing the four main column structures extending from southeast to northwest. However, both maps have angular resolution a factor $\gtrsim10$ lower than our \emph{JWST} extinction map, so the peaks are beam-diluted and fine-scale substructures are washed out.  As a result, the highest column densities in the large-beam images are significantly lower than those in our \emph{JWST} extinction map. Our extinction map reaches $N_{\rm H_2}\sim10^{23}$\,cm$^{-2}$ in the densest regions, whereas the Herschel and $^{12}$CO maps only show peak values of a few $10^{22}$\,cm$^{-2}$. While beam dilution in these coarser tracers naturally lowers the measured peak column densities, this discrepancy fundamentally arises from the different assumptions and limitations inherent to each method. Deriving $N_{\rm H_2}$ from \emph{Herschel} dust emission \citep{Molinari2010} typically assumes a single line-of-sight dust temperature; however, in dense, internally cold cores, assuming a warmer average temperature leads to a systematic underestimation of the true column density. Similarly, $N_{\rm H_2}$ derived from $^{12}$CO emission\citep{Umemoto2017} relies on a constant $X_{\rm CO}$ factor, but $^{12}$CO becomes optically thick at relatively low column densities and suffers from molecular depletion (freeze-out) in cold, dense environments, rendering it insensitive to the highest density peaks. In contrast, our NIR extinction method directly probes the total dust column, making it a more robust tracer of the densest structures.
%Our extinction map reaches $N_{\rm H_2}\sim10^{23}$\,cm$^{-2}$ in the densest regions, but the Herschel map only shows a few 10$^{22}$\,cm$^{-2}$. Beam dilution and the diffuse contribution of large-scale cloud structure lower the measured columns in these coarser tracers. 

For comparison, we constructed N-PDFs for the same region using \emph{Herschel} and $^{12}$CO data in Figures~\ref{fig:comparison}(c) and (d).  For the Herschel-derived column density map the fit yields $\sigma=0.13$ and $\alpha=3.51$, remarkably similar to our \emph{JWST} result ($\alpha=3.67$).  The nearly identical slopes indicate that Herschel dust emission traces the same dense structures that dominate the N-PDF tail. The slightly smaller $\sigma$ reflects reduced dynamic range: beam smoothing compresses observed density contrasts, thereby reducing the variance of $\ln N_{\rm H_2}$ \citep{Lombardi2015}.

In contrast, the integrated $^{12}$CO map produces a different N-PDF with $\sigma=0.25$ and $\alpha=2.45$. The broader distribution ($\sigma=0.25$) indicates that $^{12}$CO intensities span a wider relative range, partly because $^{12}$CO traces both dense and diffuse gas. The shallower tail ($\alpha=2.45$) suggests an apparent enhancement of high-density gas relative to the \emph{Herschel} and \emph{JWST} results. These differences arise from the physics of $^{12}$CO emission. Since $^{12}$CO becomes optically thick at moderate column densities ($N\gtrsim10^{21}$ cm$^{-2}$), the line saturates and no longer increases linearly with density. Consequently, $^{12}$CO emission underestimates the highest-density peaks in M16: while the line emits strongly across extended areas, it fails to brighten further in the densest clumps. This saturation effect flattens the line intensities and produces a shallower PDF tail. Furthermore, converting CO intensity to column density using a fixed $X_{\rm CO}$ factor neglects local variations in temperature and chemistry. In contrast, our dust extinction map in the NIR as measured with \emph{JWST}, maintains sensitivity to column density even in highly opaque regions.

\section{Summary} \label{sec:sum}

In this work use \emph{JWST}/NIRCam photometry to produce a $2\arcsec$-resolution extinction map of the Pillars of Creation in M16, reaching depths of $A_V\sim 100$\,mag.  The derived N-PDF shows a lognormal form at moderate extinctions ($A_V\approx$10 -- 30\,mag) transitioning to a pronounced power-law tail at higher extinctions.  Crucially, the slope of this high-extinction tail varies systematically with position: it is shallow ($\alpha\approx2.0$) in the pillar heads facing the NGC 6611 cluster, and becomes much steeper ($\alpha\approx4.0$) in regions farther from the cluster. This spatial gradient provides quantitative evidence that massive-star feedback actively compresses the exposed pillar tips, creating an enhanced fraction of dense gas consistent with radiation-driven implosion models. The fine detail of our extinction map thus reveals the dual nature of stellar feedback in M16, demonstrating how OB stars simultaneously sculpt and compress molecular clouds to regulate star formation.

\section*{Acknowledgements}

We would like to thank the anonymous referee for the very helpful comments that improved this work. We thank Prof. Biwei Jiang for helpful discussions. This work is supported by the National Natural Science Foundation of China (NSFC) 12403026, 12173034, 12322304 and 12133002, the Guangdong Basic and Applied Basic Research Foundation 2026A1515011839, the National Key R\&D program of China 2022YFA1603102. B.Q.C. acknowledges the National Natural Science Foundation of Yunnan Province 202301AV070002 and the Xingdian talent support program of Yunnan Province. H.Z. acknowledges financial support by the Chilean Government-ESO Joint Committee (Comit\'{e} Mixto ESO-Chile, No. annlang23003-es-cl). X.C. thanks to the Guangdong Province Universities and Colleges Pearl River Scholar Funded Scheme (2019).

%%%%%%%%%%%%%%%%%%%%%%%%%%%%%%%%%%%%%%%%%%%%%%%%%%
\section*{Data Availability}
 
The \emph{JWST} imaging data are available at MAST \footnote{http://dx.doi.org/10.17909/3w1e-qp71}. The extinction map of the Pillars of Creation constructed from \emph{JWST}/NIRCam is available at Zenodo \footnote{https://doi.org/10.5281/zenodo.16876647}. The photometry catalogs derived from \emph{JWST}/NIRCam will be shared on reasonable request to the corresponding author.



\bibliographystyle{mnras}
\bibliography{example} % if your bibtex file is called example.bib

@ARTICLE{Gritschneder2009,
	author = {{Gritschneder}, Matthias and {Naab}, Thorsten and {Walch}, Stefanie and {Burkert}, Andreas and {Heitsch}, Fabian},
	title = "{Driving Turbulence and Triggering Star Formation by Ionizing Radiation}",
	journal = {\apjl},
	year = 2009,
	month = mar,
	volume = {694},
	number = {1},
	pages = {L26-L30},
	doi = {10.1088/0004-637X/694/1/L26},
	archivePrefix = {arXiv},
	eprint = {0901.2113},
	primaryClass = {astro-ph.SR},
	adsurl = {https://ui.adsabs.harvard.edu/abs/2009ApJ...694L..26G}
}

@ARTICLE{Walch2012,
	author = {{Walch}, S.~K. and {Whitworth}, A.~P. and {Bisbas}, T. and {W{\"u}nsch}, R. and {Hubber}, D.},
	title = "{Dispersal of molecular clouds by ionizing radiation}",
	journal = {\mnras},
	year = 2012,
	month = nov,
	volume = {427},
	number = {1},
	pages = {625-636},
	doi = {10.1111/j.1365-2966.2012.21767.x},
	archivePrefix = {arXiv},
	eprint = {1206.6492},
	primaryClass = {astro-ph.GA},
	adsurl = {https://ui.adsabs.harvard.edu/abs/2012MNRAS.427..625W}
}

@INPROCEEDINGS{Elmegreen1998,
	author = {{Elmegreen}, B.~G.},
	title = "{Observations and Theory of Dynamical Triggers for Star Formation}",
	booktitle = {Origins},
	year = 1998,
	editor = {{Woodward}, Charles E. and {Shull}, J. Michael and {Thronson}, Jr., Harley A.},
	series = {Astronomical Society of the Pacific Conference Series},
	volume = {148},
	month = jan,
	pages = {150},
	doi = {10.48550/arXiv.astro-ph/9712352},
	archivePrefix = {arXiv},
	eprint = {astro-ph/9712352},
	primaryClass = {astro-ph},
	adsurl = {https://ui.adsabs.harvard.edu/abs/1998ASPC..148..150E}
}

@ARTICLE{Klessen2000,
	author = {{Klessen}, Ralf S.},
	title = "{One-Point Probability Distribution Functions of Supersonic Turbulent Flows in Self-gravitating Media}",
	journal = {\apj},
	year = 2000,
	month = jun,
	volume = {535},
	number = {2},
	pages = {869-886},
	doi = {10.1086/308854},
	archivePrefix = {arXiv},
	eprint = {astro-ph/0001379},
	primaryClass = {astro-ph},
	adsurl = {https://ui.adsabs.harvard.edu/abs/2000ApJ...535..869K}
}

@ARTICLE{Kainulainen2009,
	author = {{Kainulainen}, J. and {Beuther}, H. and {Henning}, T. and {Plume}, R.},
	title = "{Probing the evolution of molecular cloud structure. From quiescence to birth}",
	journal = {\aap},
	year = 2009,
	month = dec,
	volume = {508},
	number = {3},
	pages = {L35-L38},
	doi = {10.1051/0004-6361/200913605},
	archivePrefix = {arXiv},
	eprint = {0911.5648},
	primaryClass = {astro-ph.GA},
	adsurl = {https://ui.adsabs.harvard.edu/abs/2009A&A...508L..35K}
}

@ARTICLE{Federrath2013,
	author = {{Federrath}, Christoph and {Klessen}, Ralf S.},
	title = "{On the Star Formation Efficiency of Turbulent Magnetized Clouds}",
	journal = {\apj},
	year = 2013,
	month = jan,
	volume = {763},
	number = {1},
	eid = {51},
	pages = {51},
	doi = {10.1088/0004-637X/763/1/51},
	archivePrefix = {arXiv},
	eprint = {1211.6433},
	primaryClass = {astro-ph.SR},
	adsurl = {https://ui.adsabs.harvard.edu/abs/2013ApJ...763...51F}
}

@ARTICLE{Shu1977,
	author = {{Shu}, F.~H.},
	title = "{Self-similar collapse of isothermal spheres and star formation.}",
	journal = {\apj},
	year = 1977,
	month = jun,
	volume = {214},
	pages = {488-497},
	doi = {10.1086/155274},
	adsurl = {https://ui.adsabs.harvard.edu/abs/1977ApJ...214..488S}
}

@ARTICLE{Burkhart2017,
	author = {{Burkhart}, Blakesley and {Stalpes}, Kye and {Collins}, David C.},
	title = "{The Razor{\textquoteright}s Edge of Collapse: The Transition Point from Lognormal to Power-Law Distributions in Molecular Clouds}",
	journal = {\apjl},
	year = 2017,
	month = jan,
	volume = {834},
	number = {1},
	eid = {L1},
	pages = {L1},
	doi = {10.3847/2041-8213/834/1/L1},
	archivePrefix = {arXiv},
	eprint = {1609.04409},
	primaryClass = {astro-ph.GA},
	adsurl = {https://ui.adsabs.harvard.edu/abs/2017ApJ...834L...1B}
}

@ARTICLE{Tremblin2014,
	author = {{Tremblin}, P. and {Schneider}, N. and {Minier}, V. and {Didelon}, P. and {Hill}, T. and {Anderson}, L.~D. and {Motte}, F. and {Zavagno}, A. and {Andr{\'e}}, Ph. and {Arzoumanian}, D. and {Audit}, E. and {Benedettini}, M. and {Bontemps}, S. and {Csengeri}, T. and {Di Francesco}, J. and {Giannini}, T. and {Hennemann}, M. and {Nguyen Luong}, Q. and {Marston}, A.~P. and {Peretto}, N. and {Rivera-Ingraham}, A. and {Russeil}, D. and {Rygl}, K.~L.~J. and {Spinoglio}, L. and {White}, G.~J.},
	title = "{Ionization compression impact on dense gas distribution and star formation. Probability density functions around H II regions as seen by Herschel}",
	journal = {\aap},
	year = 2014,
	month = apr,
	volume = {564},
	eid = {A106},
	pages = {A106},
	doi = {10.1051/0004-6361/201322700},
	archivePrefix = {arXiv},
	eprint = {1401.7333},
	primaryClass = {astro-ph.GA},
	adsurl = {https://ui.adsabs.harvard.edu/abs/2014A&A...564A.106T}
}

@ARTICLE{Lombardi2015,
	author = {{Lombardi}, Marco and {Alves}, Jo{\~a}o and {Lada}, Charles J.},
	title = "{Molecular clouds have power-law probability distribution functions}",
	journal = {\aap},
	year = 2015,
	month = apr,
	volume = {576},
	eid = {L1},
	pages = {L1},
	doi = {10.1051/0004-6361/201525650},
	archivePrefix = {arXiv},
	eprint = {1502.03859},
	primaryClass = {astro-ph.SR},
	adsurl = {https://ui.adsabs.harvard.edu/abs/2015A&A...576L...1L}
}

@ARTICLE{Schneider2015,
	author = {{Schneider}, N. and {Ossenkopf}, V. and {Csengeri}, T. and {Klessen}, R.~S. and {Federrath}, C. and {Tremblin}, P. and {Girichidis}, P. and {Bontemps}, S. and {Andr{\'e}}, Ph.},
	title = "{Understanding star formation in molecular clouds. I. Effects of line-of-sight contamination on the column density structure}",
	journal = {\aap},
	year = 2015,
	month = mar,
	volume = {575},
	eid = {A79},
	pages = {A79},
	doi = {10.1051/0004-6361/201423569},
	archivePrefix = {arXiv},
	eprint = {1403.2996},
	primaryClass = {astro-ph.GA},
	adsurl = {https://ui.adsabs.harvard.edu/abs/2015A&A...575A..79S}
}

@ARTICLE{Li2024,
	author = {{Li}, Jun and {Chen}, Bingqiu and {Jiang}, Biwei and {Gao}, Jian and {Chen}, Xi},
	title = "{Probing the Distinct Extinction Law of the Pillars of Creation in M16 with JWST}",
	journal = {\apjl},
	year = 2024,
	month = jun,
	volume = {968},
	number = {2},
	eid = {L26},
	pages = {L26},
	doi = {10.3847/2041-8213/ad54c7},
	archivePrefix = {arXiv},
	eprint = {2406.03410},
	primaryClass = {astro-ph.GA},
	adsurl = {https://ui.adsabs.harvard.edu/abs/2024ApJ...968L..26L}
}

@MISC{Nally2023,
	author = {{Nally}, Conor},
	title = "{StarbugII: JWST PSF photometry for crowded fields}",
	howpublished = {Astrophysics Source Code Library, record ascl:2309.012},
	year = 2023,
	month = sep,
	eid = {ascl:2309.012},
	pages = {ascl:2309.012},
	archivePrefix = {ascl},
	eprint = {2309.012},
	adsurl = {https://ui.adsabs.harvard.edu/abs/2023ascl.soft09012N}
}

@ARTICLE{Girardi2005,
	author = {{Girardi}, L. and {Groenewegen}, M.~A.~T. and {Hatziminaoglou}, E. and {da Costa}, L.},
	title = "{Star counts in the Galaxy. Simulating from very deep to very shallow photometric surveys with the TRILEGAL code}",
	journal = {\aap},
	year = 2005,
	month = jun,
	volume = {436},
	number = {3},
	pages = {895-915},
	doi = {10.1051/0004-6361:20042352},
	archivePrefix = {arXiv},
	eprint = {astro-ph/0504047},
	primaryClass = {astro-ph},
	adsurl = {https://ui.adsabs.harvard.edu/abs/2005A&A...436..895G}
}

@ARTICLE{Lada1994,
	author = {{Lada}, Charles J. and {Lada}, Elizabeth A. and {Clemens}, Dan P. and {Bally}, John},
	title = "{Dust Extinction and Molecular Gas in the Dark Cloud IC 5146}",
	journal = {\apj},
	year = 1994,
	month = jul,
	volume = {429},
	pages = {694},
	doi = {10.1086/174354},
	adsurl = {https://ui.adsabs.harvard.edu/abs/1994ApJ...429..694L}
}

@ARTICLE{Schneider2022,
	author = {{Schneider}, N. and {Ossenkopf-Okada}, V. and {Clarke}, S. and {Klessen}, R.~S. and {Kabanovic}, S. and {Veltchev}, T. and {Bontemps}, S. and {Dib}, S. and {Csengeri}, T. and {Federrath}, C. and {Di Francesco}, J. and {Motte}, F. and {Andr{\'e}}, Ph. and {Arzoumanian}, D. and {Beattie}, J.~R. and {Bonne}, L. and {Didelon}, P. and {Elia}, D. and {K{\"o}nyves}, V. and {Kritsuk}, A. and {Ladjelate}, B. and {Myers}, Ph. and {Pezzuto}, S. and {Robitaille}, J.~F. and {Roy}, A. and {Seifried}, D. and {Simon}, R. and {Soler}, J. and {Ward-Thompson}, D.},
	title = "{Understanding star formation in molecular clouds. IV. Column density PDFs from quiescent to massive molecular clouds}",
	journal = {Astronomy and Astrophysics},
	year = 2022,
	month = oct,
	volume = {666},
	eid = {A165},
	pages = {A165},
	doi = {10.1051/0004-6361/202039610},
	archivePrefix = {arXiv},
	eprint = {2207.14604},
	primaryClass = {astro-ph.GA},
	adsurl = {https://ui.adsabs.harvard.edu/abs/2022A&A...666A.165S}
}

@ARTICLE{Padoan1997,
	author = {{Padoan}, Paolo and {Nordlund}, Ake and {Jones}, Bernard J.~T.},
	title = "{The universality of the stellar initial mass function}",
	journal = {Monthly Notices of the Royal Astronomical Society},
	year = 1997,
	month = jun,
	volume = {288},
	number = {1},
	pages = {145-152},
	doi = {10.1093/mnras/288.1.145},
	archivePrefix = {arXiv},
	eprint = {astro-ph/9703110},
	primaryClass = {astro-ph},
	adsurl = {https://ui.adsabs.harvard.edu/abs/1997MNRAS.288..145P}
}

@ARTICLE{Federrath2010,
	author = {{Federrath}, C. and {Roman-Duval}, J. and {Klessen}, R.~S. and {Schmidt}, W. and {Mac Low}, M. -M.},
	title = "{Comparing the statistics of interstellar turbulence in simulations and observations. Solenoidal versus compressive turbulence forcing}",
	journal = {Astronomy and Astrophysics},
	year = 2010,
	month = mar,
	volume = {512},
	eid = {A81},
	pages = {A81},
	doi = {10.1051/0004-6361/200912437},
	archivePrefix = {arXiv},
	eprint = {0905.1060},
	primaryClass = {astro-ph.SR},
	adsurl = {https://ui.adsabs.harvard.edu/abs/2010A&A...512A..81F}
}

@ARTICLE{Pound1998,
	author = {{Pound}, Marc W.},
	title = "{Molecular Gas in the Eagle Nebula}",
	journal = {The Astrophysical Journal Letters},
	year = 1998,
	month = feb,
	volume = {493},
	number = {2},
	pages = {L113-L116},
	doi = {10.1086/311131},
	adsurl = {https://ui.adsabs.harvard.edu/abs/1998ApJ...493L.113P}
}

@ARTICLE{Tremblin2013,
	author = {{Tremblin}, P. and {Minier}, V. and {Schneider}, N. and {Audit}, E. and {Hill}, T. and {Didelon}, P. and {Peretto}, N. and {Arzoumanian}, D. and {Motte}, F. and {Zavagno}, A. and {Bontemps}, S. and {Anderson}, L.~D. and {Andr{\'e}}, Ph. and {Bernard}, J.~P. and {Csengeri}, T. and {Di Francesco}, J. and {Elia}, D. and {Hennemann}, M. and {K{\"o}nyves}, V. and {Marston}, A.~P. and {Nguyen Luong}, Q. and {Rivera-Ingraham}, A. and {Roussel}, H. and {Sousbie}, T. and {Spinoglio}, L. and {White}, G.~J. and {Williams}, J.},
	title = "{Pillars and globules at the edges of H ii regions. Confronting Herschel observations and numerical simulations}",
	journal = {Astronomy and Astrophysics},
	year = 2013,
	month = dec,
	volume = {560},
	eid = {A19},
	pages = {A19},
	doi = {10.1051/0004-6361/201322233},
	archivePrefix = {arXiv},
	eprint = {1311.3664},
	primaryClass = {astro-ph.GA},
	adsurl = {https://ui.adsabs.harvard.edu/abs/2013A&A...560A..19T}
}

@ARTICLE{McCaughrean2002,
	author = {{McCaughrean}, M.~J. and {Andersen}, M.},
	title = "{The Eagle's EGGs: Fertile or sterile?}",
	journal = {Astronomy and Astrophysics},
	year = 2002,
	month = jul,
	volume = {389},
	pages = {513-518},
	doi = {10.1051/0004-6361:20020589},
	archivePrefix = {arXiv},
	eprint = {astro-ph/0202025},
	primaryClass = {astro-ph},
	adsurl = {https://ui.adsabs.harvard.edu/abs/2002A&A...389..513M}
}

@ARTICLE{Bisbas2011,
	author = {{Bisbas}, Thomas G. and {W{\"u}nsch}, Richard and {Whitworth}, Anthony P. and {Hubber}, David A. and {Walch}, Stefanie},
	title = "{Radiation-driven Implosion and Triggered Star Formation}",
	journal = {The Astrophysical Journal},
	year = 2011,
	month = aug,
	volume = {736},
	number = {2},
	eid = {142},
	pages = {142},
	doi = {10.1088/0004-637X/736/2/142},
	archivePrefix = {arXiv},
	eprint = {1105.3727},
	primaryClass = {astro-ph.SR},
	adsurl = {https://ui.adsabs.harvard.edu/abs/2011ApJ...736..142B}
}

@ARTICLE{Vazquez1994,
	author = {{Vazquez-Semadeni}, Enrique},
	title = "{Hierarchical Structure in Nearly Pressureless Flows as a Consequence of Self-similar Statistics}",
	journal = {\apj},
	year = 1994,
	month = mar,
	volume = {423},
	pages = {681},
	doi = {10.1086/173847},
	adsurl = {https://ui.adsabs.harvard.edu/abs/1994ApJ...423..681V}
}

@ARTICLE{Hill2008,
	author = {{Hill}, Alex S. and {Benjamin}, Robert A. and {Kowal}, Grzegorz and {Reynolds}, Ronald J. and {Haffner}, L. Matthew and {Lazarian}, Alex},
	title = "{The Turbulent Warm Ionized Medium: Emission Measure Distribution and MHD Simulations}",
	journal = {\apj},
	year = 2008,
	month = oct,
	volume = {686},
	number = {1},
	pages = {363-378},
	doi = {10.1086/590543},
	archivePrefix = {arXiv},
	eprint = {0805.0155},
	primaryClass = {astro-ph},
	adsurl = {https://ui.adsabs.harvard.edu/abs/2008ApJ...686..363H}
}

@ARTICLE{Kainulainen2013,
	author = {{Kainulainen}, J. and {Tan}, J.~C.},
	title = "{High-dynamic-range extinction mapping of infrared dark clouds. Dependence of density variance with sonic Mach number in molecular clouds}",
	journal = {\aap},
	year = 2013,
	month = jan,
	volume = {549},
	eid = {A53},
	pages = {A53},
	doi = {10.1051/0004-6361/201219526},
	archivePrefix = {arXiv},
	eprint = {1210.8130},
	primaryClass = {astro-ph.GA},
	adsurl = {https://ui.adsabs.harvard.edu/abs/2013A&A...549A..53K}
}

@ARTICLE{Stutz2015,
	author = {{Stutz}, A.~M. and {Kainulainen}, J.},
	title = "{Evolution of column density distributions within Orion A{\ensuremath{\star}}}",
	journal = {\aap},
	year = 2015,
	month = may,
	volume = {577},
	eid = {L6},
	pages = {L6},
	doi = {10.1051/0004-6361/201526243},
	archivePrefix = {arXiv},
	eprint = {1504.05188},
	primaryClass = {astro-ph.GA},
	adsurl = {https://ui.adsabs.harvard.edu/abs/2015A&A...577L...6S}
}

@ARTICLE{Kainulainen2011,
	author = {{Kainulainen}, J. and {Beuther}, H. and {Banerjee}, R. and {Federrath}, C. and {Henning}, T.},
	title = "{Probing the evolution of molecular cloud structure. II. From chaos to confinement}",
	journal = {\aap},
	year = 2011,
	month = jun,
	volume = {530},
	eid = {A64},
	pages = {A64},
	doi = {10.1051/0004-6361/201016383},
	archivePrefix = {arXiv},
	eprint = {1104.0678},
	primaryClass = {astro-ph.GA},
	adsurl = {https://ui.adsabs.harvard.edu/abs/2011A&A...530A..64K}
}

@ARTICLE{Schneider2013,
	author = {{Schneider}, N. and {Andr{\'e}}, Ph. and {K{\"o}nyves}, V. and {Bontemps}, S. and {Motte}, F. and {Federrath}, C. and {Ward-Thompson}, D. and {Arzoumanian}, D. and {Benedettini}, M. and {Bressert}, E. and {Didelon}, P. and {Di Francesco}, J. and {Griffin}, M. and {Hennemann}, M. and {Hill}, T. and {Palmeirim}, P. and {Pezzuto}, S. and {Peretto}, N. and {Roy}, A. and {Rygl}, K.~L.~J. and {Spinoglio}, L. and {White}, G.},
	title = "{What Determines the Density Structure of Molecular Clouds? A Case Study of Orion B with Herschel}",
	journal = {\apjl},
	year = 2013,
	month = apr,
	volume = {766},
	number = {2},
	eid = {L17},
	pages = {L17},
	doi = {10.1088/2041-8205/766/2/L17},
	archivePrefix = {arXiv},
	eprint = {1304.0327},
	primaryClass = {astro-ph.GA},
	adsurl = {https://ui.adsabs.harvard.edu/abs/2013ApJ...766L..17S}
}

@ARTICLE{Linsky2007,
	author = {{Linsky}, Jeffrey L. and {Gagn{\'e}}, Marc and {Mytyk}, Anna and {McCaughrean}, Mark and {Andersen}, Morten},
	title = "{Chandra Observations of the Eagle Nebula. I. Embedded Young Stellar Objects near the Pillars of Creation}",
	journal = {\apj},
	year = 2007,
	month = jan,
	volume = {654},
	number = {1},
	pages = {347-360},
	doi = {10.1086/508763},
	archivePrefix = {arXiv},
	eprint = {astro-ph/0610279},
	primaryClass = {astro-ph},
	adsurl = {https://ui.adsabs.harvard.edu/abs/2007ApJ...654..347L}
}

@ARTICLE{Kuhn2019,
	author = {{Kuhn}, Michael A. and {Hillenbrand}, Lynne A. and {Sills}, Alison and {Feigelson}, Eric D. and {Getman}, Konstantin V.},
	title = "{Kinematics in Young Star Clusters and Associations with Gaia DR2}",
	journal = {The Astrophysical Journal},
	year = 2019,
	month = jan,
	volume = {870},
	number = {1},
	eid = {32},
	pages = {32},
	doi = {10.3847/1538-4357/aaef8c},
	archivePrefix = {arXiv},
	eprint = {1807.02115},
	primaryClass = {astro-ph.GA},
	adsurl = {https://ui.adsabs.harvard.edu/abs/2019ApJ...870...32K}
}

@ARTICLE{Dufton2006,
	author = {{Dufton}, P.~L. and {Smartt}, S.~J. and {Lee}, J.~K. and {Ryans}, R.~S.~I. and {Hunter}, I. and {Evans}, C.~J. and {Herrero}, A. and {Trundle}, C. and {Lennon}, D.~J. and {Irwin}, M.~J. and {Kaufer}, A.},
	title = "{The VLT-FLAMES survey of massive stars: stellar parameters and rotational velocities in NGC 3293, NGC 4755 and NGC 6611}",
	journal = {Astronomy and Astrophysics},
	year = 2006,
	month = oct,
	volume = {457},
	number = {1},
	pages = {265-280},
	doi = {10.1051/0004-6361:20065392},
	archivePrefix = {arXiv},
	eprint = {astro-ph/0606409},
	primaryClass = {astro-ph},
	adsurl = {https://ui.adsabs.harvard.edu/abs/2006A&A...457..265D}
}

@ARTICLE{Hester1996,
	author = {{Hester}, J.~J. and {Scowen}, P.~A. and {Sankrit}, R. and {Lauer}, T.~R. and {Ajhar}, E.~A. and {Baum}, W.~A. and {Code}, A. and {Currie}, D.~G. and {Danielson}, G.~E. and {Ewald}, S.~P. and {Faber}, S.~M. and {Grillmair}, C.~J. and {Groth}, E.~J. and {Holtzman}, J.~A. and {Hunter}, D.~A. and {Kristian}, J. and {Light}, R.~M. and {Lynds}, C.~R. and {Monet}, D.~G. and {O'Neil}, Jr., E.~J. and {Shaya}, E.~J. and {Seidelmann}, P.~K. and {Westphal}, J.~A.},
	title = "{Hubble Space Telescope WFPC2 Imaging of M16: Photoevaporation and Emerging Young Stellar Objects}",
	journal = {The Astronomical Journal},
	year = 1996,
	month = jun,
	volume = {111},
	pages = {2349},
	doi = {10.1086/117968},
	adsurl = {https://ui.adsabs.harvard.edu/abs/1996AJ....111.2349H}
}

@ARTICLE{Indebetouw2007,
	author = {{Indebetouw}, R. and {Robitaille}, T.~P. and {Whitney}, B.~A. and {Churchwell}, E. and {Babler}, B. and {Meade}, M. and {Watson}, C. and {Wolfire}, M.},
	title = "{Embedded Star Formation in the Eagle Nebula with Spitzer GLIMPSE}",
	journal = {\apj},
	year = 2007,
	month = sep,
	volume = {666},
	number = {1},
	pages = {321-338},
	doi = {10.1086/520316},
	archivePrefix = {arXiv},
	eprint = {0707.1895},
	primaryClass = {astro-ph},
	adsurl = {https://ui.adsabs.harvard.edu/abs/2007ApJ...666..321I}
}

@ARTICLE{Guarcello2010,
	author = {{Guarcello}, M.~G. and {Micela}, G. and {Peres}, G. and {Prisinzano}, L. and {Sciortino}, S.},
	title = "{Chronology of star formation and disk evolution in the Eagle Nebula}",
	journal = {\aap},
	year = 2010,
	month = oct,
	volume = {521},
	eid = {A61},
	pages = {A61},
	doi = {10.1051/0004-6361/201014351},
	archivePrefix = {arXiv},
	eprint = {1008.0422},
	primaryClass = {astro-ph.SR},
	adsurl = {https://ui.adsabs.harvard.edu/abs/2010A&A...521A..61G}
}

@ARTICLE{White1999,
	author = {{White}, G.~J. and {Nelson}, R.~P. and {Holland}, W.~S. and {Robson}, E.~I. and {Greaves}, J.~S. and {McCaughrean}, M.~J. and {Pilbratt}, G.~L. and {Balser}, D.~S. and {Oka}, T. and {Sakamoto}, S. and {Hasegawa}, T. and {McCutcheon}, W.~H. and {Matthews}, H.~E. and {Fridlund}, C.~V.~M. and {Tothill}, N.~F.~H. and {Huldtgren}, M. and {Deane}, J.~R.},
	title = "{The Eagle Nebula's fingers - pointers to the earliest stages of star formation?}",
	journal = {\aap},
	year = 1999,
	month = feb,
	volume = {342},
	pages = {233-256},
	adsurl = {https://ui.adsabs.harvard.edu/abs/1999A&A...342..233W}
}

@ARTICLE{Hill2012,
	author = {{Hill}, T. and {Motte}, F. and {Didelon}, P. and {White}, G.~J. and {Marston}, A.~P. and {Nguy{\^e}n Luong}, Q. and {Bontemps}, S. and {Andr{\'e}}, Ph. and {Schneider}, N. and {Hennemann}, M. and {Sauvage}, M. and {Di Francesco}, J. and {Minier}, V. and {Anderson}, L.~D. and {Bernard}, J.~P. and {Elia}, D. and {Griffin}, M.~J. and {Li}, J.~Z. and {Peretto}, N. and {Pezzuto}, S. and {Polychroni}, D. and {Roussel}, H. and {Rygl}, K.~L.~J. and {Schisano}, E. and {Sousbie}, T. and {Testi}, L. and {Thompson}, D. Ward and {Zavagno}, A.},
	title = "{The M 16 molecular complex under the influence of NGC 6611. Herschel's perspective of the heating effect on the Eagle Nebula}",
	journal = {\aap},
	year = 2012,
	month = jun,
	volume = {542},
	eid = {A114},
	pages = {A114},
	doi = {10.1051/0004-6361/201219009},
	archivePrefix = {arXiv},
	eprint = {1204.6317},
	primaryClass = {astro-ph.SR},
	adsurl = {https://ui.adsabs.harvard.edu/abs/2012A&A...542A.114H}
}

@ARTICLE{Hensley2023,
	author = {{Hensley}, Brandon S. and {Draine}, B.~T.},
	title = "{The Astrodust+PAH Model: A Unified Description of the Extinction, Emission, and Polarization from Dust in the Diffuse Interstellar Medium}",
	journal = {\apj},
	year = 2023,
	month = may,
	volume = {948},
	number = {1},
	eid = {55},
	pages = {55},
	doi = {10.3847/1538-4357/acc4c2},
	archivePrefix = {arXiv},
	eprint = {2208.12365},
	primaryClass = {astro-ph.GA},
	adsurl = {https://ui.adsabs.harvard.edu/abs/2023ApJ...948...55H}
}

@ARTICLE{Ward2014,
	author = {{Ward}, Rachel L. and {Wadsley}, James and {Sills}, Alison},
	title = "{Evolving molecular cloud structure and the column density probability distribution function}",
	journal = {\mnras},
	year = 2014,
	month = dec,
	volume = {445},
	number = {2},
	pages = {1575-1583},
	doi = {10.1093/mnras/stu1868},
	archivePrefix = {arXiv},
	eprint = {1409.2540},
	primaryClass = {astro-ph.GA},
	adsurl = {https://ui.adsabs.harvard.edu/abs/2014MNRAS.445.1575W}
}

@ARTICLE{Karim2023,
	author = {{Karim}, Ramsey L. and {Pound}, Marc W. and {Tielens}, Alexander G.~G.~M. and {Tiwari}, Maitraiyee and {Bonne}, Lars and {Wolfire}, Mark G. and {Schneider}, Nicola and {Kavak}, {\"U}mit and {Mundy}, Lee G. and {Simon}, Robert and {G{\"u}sten}, Rolf and {Stutzki}, J{\"u}rgen and {Wyrowski}, Friedrich and {Honingh}, Netty},
	title = "{SOFIA FEEDBACK Survey: The Pillars of Creation in [C II] and Molecular Lines}",
	journal = {\aj},
	year = 2023,
	month = dec,
	volume = {166},
	number = {6},
	eid = {240},
	pages = {240},
	doi = {10.3847/1538-3881/acff6c},
	archivePrefix = {arXiv},
	eprint = {2309.14637},
	primaryClass = {astro-ph.GA},
	adsurl = {https://ui.adsabs.harvard.edu/abs/2023AJ....166..240K}
}

@ARTICLE{Dewangan2024,
	author = {{Dewangan}, L.~K. and {Jadhav}, O.~R. and {Maity}, A.~K. and {Bhadari}, N.~K. and {Sharma}, Saurabh and {Padovani}, M. and {Baug}, T. and {Mayya}, Y.~D. and {Pandey}, Rakesh},
	title = "{Deciphering the hidden structures of HH 216 and Pillar IV in M16: results from JWST and HST}",
	journal = {\mnras},
	year = 2024,
	month = mar,
	volume = {528},
	number = {3},
	pages = {3909-3926},
	doi = {10.1093/mnras/stae150},
	archivePrefix = {arXiv},
	eprint = {2401.06016},
	primaryClass = {astro-ph.GA},
	adsurl = {https://ui.adsabs.harvard.edu/abs/2024MNRAS.528.3909D}
}

@ARTICLE{Umemoto2017,
	author = {{Umemoto}, Tomofumi and {Minamidani}, Tetsuhiro and {Kuno}, Nario and {Fujita}, Shinji and {Matsuo}, Mitsuhiro and {Nishimura}, Atsushi and {Torii}, Kazufumi and {Tosaki}, Tomoka and {Kohno}, Mikito and {Kuriki}, Mika and {Tsuda}, Yuya and {Hirota}, Akihiko and {Ohashi}, Satoshi and {Yamagishi}, Mitsuyoshi and {Handa}, Toshihiro and {Nakanishi}, Hiroyuki and {Omodaka}, Toshihiro and {Koide}, Nagito and {Matsumoto}, Naoko and {Onishi}, Toshikazu and {Tokuda}, Kazuki and {Seta}, Masumichi and {Kobayashi}, Yukinori and {Tachihara}, Kengo and {Sano}, Hidetoshi and {Hattori}, Yusuke and {Onodera}, Sachiko and {Oasa}, Yumiko and {Kamegai}, Kazuhisa and {Tsuboi}, Masato and {Sofue}, Yoshiaki and {Higuchi}, Aya E. and {Chibueze}, James O. and {Mizuno}, Norikazu and {Honma}, Mareki and {Muller}, Erik and {Inoue}, Tsuyoshi and {Morokuma-Matsui}, Kana and {Shinnaga}, Hiroko and {Ozawa}, Takeaki and {Takahashi}, Ryo and {Yoshiike}, Satoshi and {Costes}, Jean and {Kuwahara}, Sho},
	title = "{FOREST unbiased Galactic plane imaging survey with the Nobeyama 45 m telescope (FUGIN). I. Project overview and initial results}",
	journal = {\pasj},
	year = 2017,
	month = oct,
	volume = {69},
	number = {5},
	eid = {78},
	pages = {78},
	doi = {10.1093/pasj/psx061},
	archivePrefix = {arXiv},
	eprint = {1707.05981},
	primaryClass = {astro-ph.GA},
	adsurl = {https://ui.adsabs.harvard.edu/abs/2017PASJ...69...78U}
}

@ARTICLE{Molinari2010,
	author = {{Molinari}, S. and {Swinyard}, B. and {Bally}, J. and {Barlow}, M. and {Bernard}, J. -P. and {Martin}, P. and {Moore}, T. and {Noriega-Crespo}, A. and {Plume}, R. and {Testi}, L. and {Zavagno}, A. and {Abergel}, A. and {Ali}, B. and {Andr{\'e}}, P. and {Baluteau}, J. -P. and {Benedettini}, M. and {Bern{\'e}}, O. and {Billot}, N.~P. and {Blommaert}, J. and {Bontemps}, S. and {Boulanger}, F. and {Brand}, J. and {Brunt}, C. and {Burton}, M. and {Campeggio}, L. and {Carey}, S. and {Caselli}, P. and {Cesaroni}, R. and {Cernicharo}, J. and {Chakrabarti}, S. and {Chrysostomou}, A. and {Codella}, C. and {Cohen}, M. and {Compiegne}, M. and {Davis}, C.~J. and {de Bernardis}, P. and {de Gasperis}, G. and {Di Francesco}, J. and {di Giorgio}, A.~M. and {Elia}, D. and {Faustini}, F. and {Fischera}, J.~F. and {Fukui}, Y. and {Fuller}, G.~A. and {Ganga}, K. and {Garcia-Lario}, P. and {Giard}, M. and {Giardino}, G. and {Glenn}, J. and {Goldsmith}, P. and {Griffin}, M. and {Hoare}, M. and {Huang}, M. and {Jiang}, B. and {Joblin}, C. and {Joncas}, G. and {Juvela}, M. and {Kirk}, J. and {Lagache}, G. and {Li}, J.~Z. and {Lim}, T.~L. and {Lord}, S.~D. and {Lucas}, P.~W. and {Maiolo}, B. and {Marengo}, M. and {Marshall}, D. and {Masi}, S. and {Massi}, F. and {Matsuura}, M. and {Meny}, C. and {Minier}, V. and {Miville-Desch{\^e}nes}, M. -A. and {Montier}, L. and {Motte}, F. and {M{\"u}ller}, T.~G. and {Natoli}, P. and {Neves}, J. and {Olmi}, L. and {Paladini}, R. and {Paradis}, D. and {Pestalozzi}, M. and {Pezzuto}, S. and {Piacentini}, F. and {Pomar{\`e}s}, M. and {Popescu}, C.~C. and {Reach}, W.~T. and {Richer}, J. and {Ristorcelli}, I. and {Roy}, A. and {Royer}, P. and {Russeil}, D. and {Saraceno}, P. and {Sauvage}, M. and {Schilke}, P. and {Schneider-Bontemps}, N. and {Schuller}, F. and {Schultz}, B. and {Shepherd}, D.~S. and {Sibthorpe}, B. and {Smith}, H.~A. and {Smith}, M.~D. and {Spinoglio}, L. and {Stamatellos}, D. and {Strafella}, F. and {Stringfellow}, G. and {Sturm}, E. and {Taylor}, R. and {Thompson}, M.~A. and {Tuffs}, R.~J. and {Umana}, G. and {Valenziano}, L. and {Vavrek}, R. and {Viti}, S. and {Waelkens}, C. and {Ward-Thompson}, D. and {White}, G. and {Wyrowski}, F. and {Yorke}, H.~W. and {Zhang}, Q.},
	title = "{Hi-GAL: The Herschel Infrared Galactic Plane Survey}",
	journal = {\pasp},
	year = 2010,
	month = mar,
	volume = {122},
	number = {889},
	pages = {314},
	doi = {10.1086/651314},
	archivePrefix = {arXiv},
	eprint = {1001.2106},
	primaryClass = {astro-ph.GA},
	adsurl = {https://ui.adsabs.harvard.edu/abs/2010PASP..122..314M}
}

@ARTICLE{Kainulainen2014,
	author = {{Kainulainen}, Jouni and {Federrath}, Christoph and {Henning}, Thomas},
	title = "{Unfolding the Laws of Star Formation: The Density Distribution of Molecular Clouds}",
	journal = {Science},
	year = 2014,
	month = apr,
	volume = {344},
	number = {6180},
	pages = {183-185},
	doi = {10.1126/science.1248724},
	archivePrefix = {arXiv},
	eprint = {1404.2722},
	primaryClass = {astro-ph.GA},
	adsurl = {https://ui.adsabs.harvard.edu/abs/2014Sci...344..183K}
}

@ARTICLE{Sugitani2002,
	author = {{Sugitani}, K. and {Tamura}, M. and {Nakajima}, Y. and {Nagashima}, C. and {Nagayama}, T. and {Nakaya}, H. and {Pickles}, A.~J. and {Nagata}, T. and {Sato}, S. and {Fukuda}, N. and {Ogura}, K.},
	title = "{Near-Infrared Study of M16: Star Formation in the Elephant Trunks}",
	journal = {\apjl},
	year = 2002,
	month = jan,
	volume = {565},
	number = {1},
	pages = {L25-L28},
	doi = {10.1086/339196},
	adsurl = {https://ui.adsabs.harvard.edu/abs/2002ApJ...565L..25S}
}

@ARTICLE{Hillenbrand1993,
	author = {{Hillenbrand}, Lynne A. and {Massey}, Philip and {Strom}, Stephen E. and {Merrill}, K. Michael},
	title = "{NGC 6611: A Cluster Caught in the Act}",
	journal = {\aj},
	year = 1993,
	month = nov,
	volume = {106},
	pages = {1906},
	doi = {10.1086/116774},
	adsurl = {https://ui.adsabs.harvard.edu/abs/1993AJ....106.1906H}
}

@ARTICLE{Burkhart2012,
	author = {{Burkhart}, Blakesley and {Lazarian}, A.},
	title = "{The Column Density Variance-\{\textbackslashcal M\}\_s Relationship}",
	journal = {\apjl},
	year = 2012,
	month = aug,
	volume = {755},
	number = {1},
	eid = {L19},
	pages = {L19},
	doi = {10.1088/2041-8205/755/1/L19},
	archivePrefix = {arXiv},
	eprint = {1205.3792},
	primaryClass = {astro-ph.GA},
	adsurl = {https://ui.adsabs.harvard.edu/abs/2012ApJ...755L..19B}
}

@ARTICLE{Bolatto2013,
	author = {{Bolatto}, Alberto D. and {Wolfire}, Mark and {Leroy}, Adam K.},
	title = "{The CO-to-H$_{2}$ Conversion Factor}",
	journal = {\araa},
	year = 2013,
	month = aug,
	volume = {51},
	number = {1},
	pages = {207-268},
	doi = {10.1146/annurev-astro-082812-140944},
	archivePrefix = {arXiv},
	eprint = {1301.3498},
	primaryClass = {astro-ph.GA},
	adsurl = {https://ui.adsabs.harvard.edu/abs/2013ARA&A..51..207B}
}

@ARTICLE{Kritsuk2011,
	author = {{Kritsuk}, Alexei G. and {Norman}, Michael L. and {Wagner}, Rick},
	title = "{On the Density Distribution in Star-forming Interstellar Clouds}",
	journal = {\apjl},
	year = 2011,
	month = jan,
	volume = {727},
	number = {1},
	eid = {L20},
	pages = {L20},
	doi = {10.1088/2041-8205/727/1/L20},
	archivePrefix = {arXiv},
	eprint = {1007.2950},
	primaryClass = {astro-ph.GA},
	adsurl = {https://ui.adsabs.harvard.edu/abs/2011ApJ...727L..20K}
}

@ARTICLE{Collins2012,
	author = {{Collins}, David C. and {Kritsuk}, Alexei G. and {Padoan}, Paolo and {Li}, Hui and {Xu}, Hao and {Ustyugov}, Sergey D. and {Norman}, Michael L.},
	title = "{The Two States of Star-forming Clouds}",
	journal = {\apj},
	year = 2012,
	month = may,
	volume = {750},
	number = {1},
	eid = {13},
	pages = {13},
	doi = {10.1088/0004-637X/750/1/13},
	archivePrefix = {arXiv},
	eprint = {1202.2594},
	primaryClass = {astro-ph.SR},
	adsurl = {https://ui.adsabs.harvard.edu/abs/2012ApJ...750...13C}
}

@ARTICLE{Chen2015,
	author = {{Chen}, B. -Q. and {Liu}, X. -W. and {Yuan}, H. -B. and {Huang}, Y. and {Xiang}, M. -S.},
	title = "{Dust-to-gas ratio, X$_{CO}$ factor and CO-dark gas in the Galactic anticentre: an observational study}",
	journal = {\mnras},
	year = 2015,
	month = apr,
	volume = {448},
	number = {3},
	pages = {2187-2196},
	doi = {10.1093/mnras/stv103},
	archivePrefix = {arXiv},
	eprint = {1501.03606},
	primaryClass = {astro-ph.GA},
	adsurl = {https://ui.adsabs.harvard.edu/abs/2015MNRAS.448.2187C}
}

@ARTICLE{Chen2018,
	author = {{Chen}, Hope How-Huan and {Burkhart}, Blakesley and {Goodman}, Alyssa and {Collins}, David C.},
	title = "{The Anatomy of the Column Density Probability Distribution Function (N-PDF)}",
	journal = {\apj},
	year = 2018,
	month = jun,
	volume = {859},
	number = {2},
	eid = {162},
	pages = {162},
	doi = {10.3847/1538-4357/aabaf6},
	archivePrefix = {arXiv},
	eprint = {1707.09356},
	primaryClass = {astro-ph.GA},
	adsurl = {https://ui.adsabs.harvard.edu/abs/2018ApJ...859..162C}
}

@ARTICLE{Schuller2006,
	author = {{Schuller}, F. and {Leurini}, S. and {Hieret}, C. and {Menten}, K.~M. and {Philipp}, S.~D. and {G{\"u}sten}, R. and {Schilke}, P. and {Nyman}, L. -{\r{A}}.},
	title = "{Molecular excitation in the Eagle nebula's fingers}",
	journal = {\aap},
	year = 2006,
	month = aug,
	volume = {454},
	number = {2},
	pages = {L87-L90},
	doi = {10.1051/0004-6361:20065510},
	archivePrefix = {arXiv},
	eprint = {astro-ph/0606155},
	primaryClass = {astro-ph},
	adsurl = {https://ui.adsabs.harvard.edu/abs/2006A&A...454L..87S}
}

@ARTICLE{Wen2025,
	author = {{Wen}, Jing and {Chen}, Bingqiu and {Gao}, Jian and {Li}, Jun and {Yang}, Ming and {Jiang}, Biwei},
	title = "{Evidence of triggered star formation in the Pillars of Creation from JWST observations}",
	journal = {Nature Astronomy},
	year = 2025,
	month = dec,
	volume = {9},
	pages = {1845-1853},
	doi = {10.1038/s41550-025-02683-8},
	adsurl = {https://ui.adsabs.harvard.edu/abs/2025NatAs...9.1845W}
}

@ARTICLE{Gramze2025,
	author = {{Gramze}, Savannah and {Ginsburg}, Adam and {Budaiev}, Nazar and {Bulatek}, Alyssa and {Richardson}, Theo and {Barnes}, A.~T. and {Santa-Maria}, Miriam G. and {Sormani}, Mattia C. and {Lu}, Xing and {Nogueras-Lara}, Francisco and {Gaches}, Brandt A.~L. and {Battersby}, Cara D. and {Wallace}, Jennifer and {Walker}, Daniel L. and {Mills}, Elisabeth A.~C. and {Mattern}, Michael},
	title = "{Mapping CO Ice in a Star-Forming Filament in the 3 kpc Arm with JWST}",
	journal = {arXiv e-prints},
	year = 2025,
	month = sep,
	eid = {arXiv:2509.21763},
	pages = {arXiv:2509.21763},
	doi = {10.48550/arXiv.2509.21763},
	archivePrefix = {arXiv},
	eprint = {2509.21763},
	primaryClass = {astro-ph.GA},
	adsurl = {https://ui.adsabs.harvard.edu/abs/2025arXiv250921763G}
}

@ARTICLE{Ginsburg2023,
	author = {{Ginsburg}, Adam and {Barnes}, Ashley T. and {Battersby}, Cara D. and {Bulatek}, Alyssa and {Gramze}, Savannah and {Henshaw}, Jonathan D. and {Jeff}, Desmond and {Lu}, Xing and {Mills}, E.~A.~C. and {Walker}, Daniel L.},
	title = "{JWST Reveals Widespread CO Ice and Gas Absorption in the Galactic Center Cloud G0.253+0.016}",
	journal = {\apj},
	year = 2023,
	month = dec,
	volume = {959},
	number = {1},
	eid = {36},
	pages = {36},
	doi = {10.3847/1538-4357/acfc34},
	archivePrefix = {arXiv},
	eprint = {2308.16050},
	primaryClass = {astro-ph.GA},
	adsurl = {https://ui.adsabs.harvard.edu/abs/2023ApJ...959...36G}
}

@ARTICLE{Bravo2025,
	author = {{Bravo Ferres}, Luc{\'\i}a and {Nogueras-Lara}, Francisco and {Sch{\"o}del}, Rainer and {Fedriani}, Rub{\'e}n and {Ginsburg}, Adam and {Crowe}, Samuel and {Tan}, Jonathan C. and {Andersen}, Morten and {Armstrong}, Joseph and {Cheng}, Yu and {Li}, Zhi-Yun},
	title = "{The JWST-NIRCam view of Sagittarius C: III. The extinction curve}",
	journal = {\aap},
	year = 2025,
	month = dec,
	volume = {704},
	eid = {A130},
	pages = {A130},
	doi = {10.1051/0004-6361/202556490},
	archivePrefix = {arXiv},
	eprint = {2510.10749},
	primaryClass = {astro-ph.GA},
	adsurl = {https://ui.adsabs.harvard.edu/abs/2025A&A...704A.130B}
}

@ARTICLE{Fahrion2023,
	author = {{Fahrion}, Katja and {De Marchi}, Guido},
	title = "{Extending the extinction law in 30 Doradus to the infrared with JWST}",
	journal = {\aap},
	year = 2023,
	month = mar,
	volume = {671},
	eid = {L14},
	pages = {L14},
	doi = {10.1051/0004-6361/202346240},
	archivePrefix = {arXiv},
	eprint = {2303.04820},
	primaryClass = {astro-ph.GA},
	adsurl = {https://ui.adsabs.harvard.edu/abs/2023A&A...671L..14F}
}

@ARTICLE{Vermot2025,
	author = {{Vermot}, Pierre},
	title = "{JWST/NIRCam: Stellar Extinction Reveals the CND of NGC 1068}",
	journal = {Research Notes of the American Astronomical Society},
	year = 2025,
	month = dec,
	volume = {9},
	number = {12},
	eid = {331},
	pages = {331},
	doi = {10.3847/2515-5172/ae2a2b},
	adsurl = {https://ui.adsabs.harvard.edu/abs/2025RNAAS...9..331V}
}

%\begin{thebibliography}{99}
%\end{thebibliography}

%%%%%%%%%%%%%%%%%%%%%%%%%%%%%%%%%%%%%%%%%%%%%%%%%%

%%%%%%%%%%%%%%%%% APPENDICES %%%%%%%%%%%%%%%%%%%%%

\appendix
\section{Spatial Distribution and Extinction of Background Sources}\label{sec:appendixA}
\begin{figure*}
	\centering
	\includegraphics[width=0.5\textwidth]{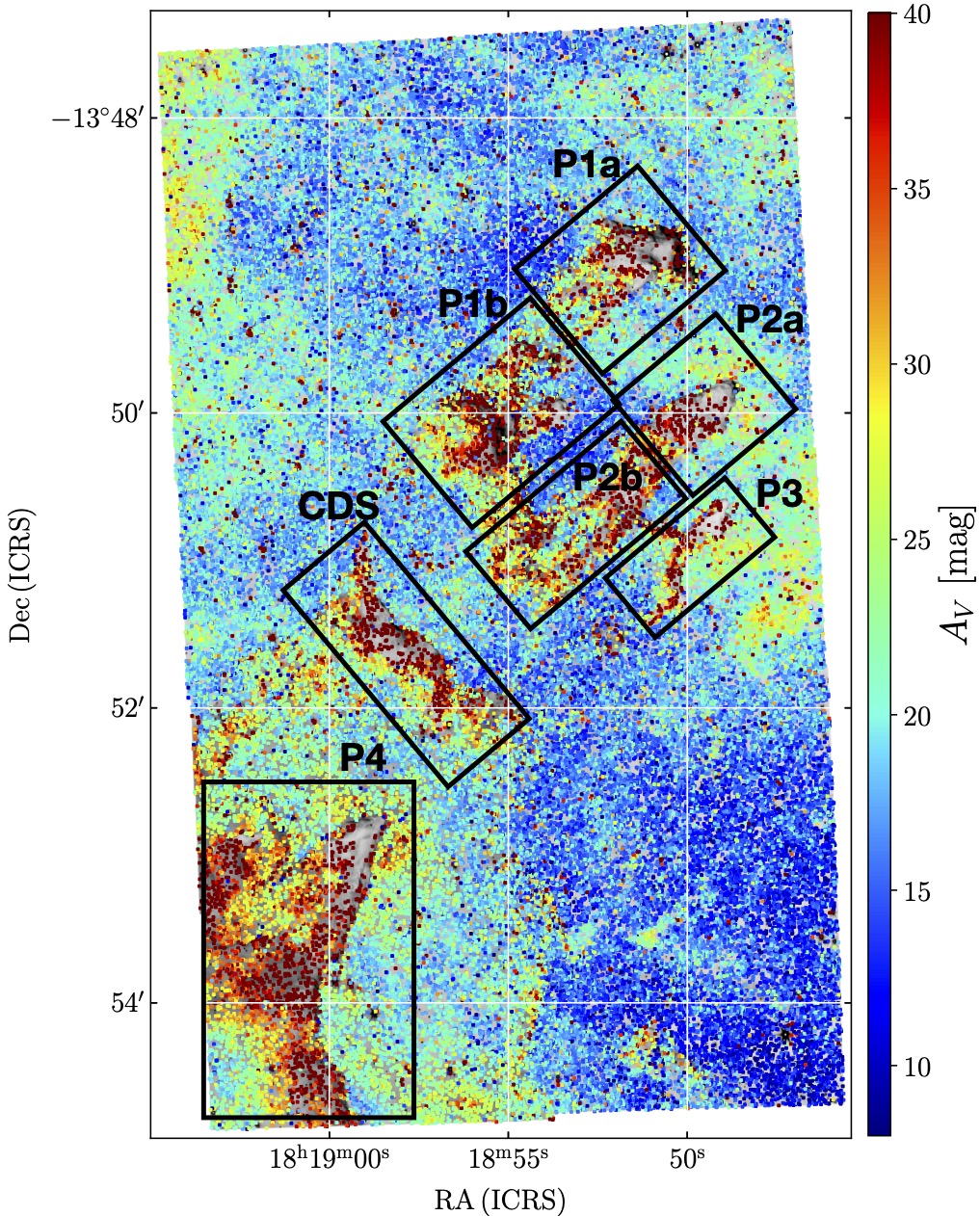}
	\caption{Spatial distribution of the background field stars used to construct the continuous extinction map, overlaid on the F335M band image. The sample consists of $\sim 10^5$ stars across the observed field. Each point represents an individual star, color-coded by its derived visual extinction ($A_V$). The rectangular boxes with labels mark the limits of the subregions used for the N-PDF fitting in Figure \ref{fig:npdf_subregions}. The high spatial density of these probing stars, characterized by a median nearest-neighbour distance of 0.51\,\arcsec, provides sufficient spatial sampling to justify the 2\,\arcsec\ resolution of the smoothed extinction map presented in Figure  \ref{fig:ce_map}. }
	\label{fig:appendix}
\end{figure*}

\section{Probability distribution functions}\label{sec:npdf_fit}
We analyze the N-PDF of M16 by computing the histogram of the logarithmic column density, $\eta = \ln(N/\langle N\rangle)$, where $N$ is traced by the $A_V$ extinction and $\langle N\rangle$ is mean column density. The N-PDF is modeled as a combination of a log-normal part and a high-density power-law tail. Specifically, the lognormal component is written as:
\begin{equation}
p(\eta) = \frac{p_0}{\sqrt{2\pi}\sigma_\eta}\exp\left[-\frac{(\eta-\mu)^2}{2\sigma^2}\right],
\end{equation}
where $\sigma$ is the dispersion, $\mu$ is the mean logarithmic column density, and $p_0$ is a normalization constant ensuring the PDF integrates to unity \citep[e.g.][]{Vazquez1994,Federrath2013}. The width $\sigma$ encodes the dispersion of the density field set by turbulence \citep{Federrath2010,Burkhart2017}: larger $\sigma$ corresponds to stronger turbulent compressions. In purely turbulence-dominated clouds, the N-PDF remains close to lognormal.
		
At high column densities, self-gravity produces an excess above the lognormal part. We fit this high-$\eta$ regime with a power-law form:
\begin{equation}
p(\eta) \propto \exp(-\alpha\eta),,
\end{equation}
where $\alpha$ is the power-law index. Equivalently, this corresponds to $p(N)\propto N^{-\alpha}$ \citep{Kritsuk2011,Collins2012,Federrath2013}. A shallow tail with small $\alpha$ value indicates a relatively large fraction of very dense gas, as expected in regions with active gravitational collapse \citep[e.g.][]{Kainulainen2009,Stutz2015}.
		
We bin the measured $\eta$ values (using logarithmic bins) and estimate the uncertainty in each bin from Poisson statistics. Following the methodological approaches of previous studies \citep[e.g.,][]{Stutz2015, Schneider2015}, we carefully define the fitting ranges for the lognormal and power-law functions. To ensure statistical robustness, both the lower and upper bounds of the overall N-PDF are truncated to avoid extreme density regimes dominated by low-number pixel statistics. We then fit the lognormal and power-law functions over defined ranges. Specifically, the upper bound of the log-normal range is defined as the point where the observed N-PDF exceeds the extrapolated log-normal model by more than three times the statistical noise in $p(\eta)$. For the power-law tail, the definition of the fitting regime is driven by the requirement of excluding areas with curvature in the N-PDFs, thus allowing for a transitional regime between the two fits. The fitting ranges and best-fit parameters ($\sigma$ for the lognormal and $\alpha$ for the power-law tail) are obtained by minimizing chi-square. The resulting fitting ranges for both components are now explicitly shaded in Figure \ref{fig:npdf_subregions}, and the corresponding parameters are summarized in Table \ref{tab:fit_res}.

%We bin the measured $\eta$ values (using logarithmic bins) and estimate the uncertainty in each bin from Poisson statistics. We then fit the lognormal and power-law functions over appropriate ranges. The lognormal fit is applied to the low-- and intermediate--$\eta$ part of the N-PDF up to the point where the PDF begins to deviate upward, while the power-law is fit only to the highest-$\eta$ portion where the N-PDF is roughly linear on a log-log scale. These fitting intervals are chosen by inspecting the N-PDF shape to exclude the turnover between regimes \citep[following e.g.][]{Kainulainen2009,Stutz2015,Lombardi2015}. The fitting ranges and best-fit parameters ($\sigma$ for the lognormal and $\alpha$ for the tail) are obtained by minimizing chi-square and are listed in Table \ref{tab:fit_res}. 

%%%%%%%%%%%%%%%%%%%%%%%%%%%%%%%%%%%%%%%%%%%%%%%%%%

% Don't change these lines
\bsp	% typesetting comment
\label{lastpage}
\end{document}